\pdfoutput=1

\documentclass[12pt]{iopart}

\usepackage[T1]{fontenc}
\usepackage[utf8]{inputenc}

\usepackage[usenames, dvipsnames]{xcolor}
\definecolor{linkcolor}{rgb}{0.0,0.3,0.5}
\usepackage[unicode,colorlinks=true,linkcolor=linkcolor,urlcolor=linkcolor,citecolor=linkcolor,pdfusetitle]{hyperref}
\usepackage{cite}
\usepackage[all]{hypcap}
\usepackage{graphicx}
\usepackage{xspace}
\usepackage{mathrsfs}
\usepackage{pifont} %
\usepackage{lmodern}

\usepackage{iopams}

\expandafter\let\csname equation*\endcsname\relax
\expandafter\let\csname endequation*\endcsname\relax

\usepackage{amsmath,amsfonts,amssymb}
\usepackage{relsize, exscale}

\usepackage{longtable}

\usepackage{microtype}
\usepackage[normalem]{ulem} %

\graphicspath{{../Scripts/}{../Scripts/LIGOcomp/}}

\newcommand{\figCOMuncorrected}{%
\begin{figure}
  \includegraphics[width = 6in]{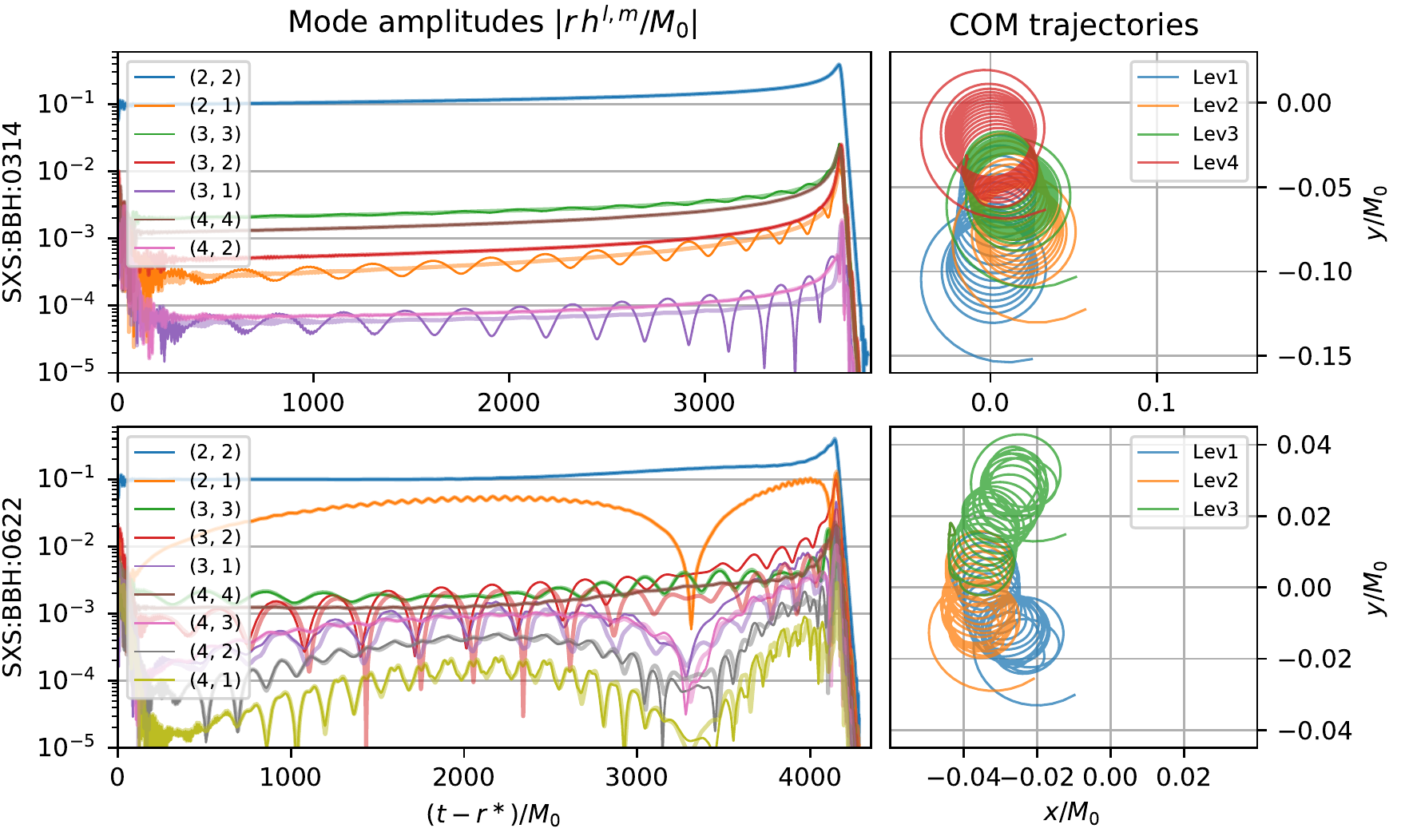}
  \caption{\label{fig:COMuncorrected} Center of Mass (COM) corrected and uncorrected waveform mode
amplitudes (left) and COM drift in simulation units (right) for spin-aligned system SXS:BBH:0314 (top)
and precessing system SXS:BBH:0622 (bottom). For the waveform mode amplitude plots on the left, the thick
translucent curves show the COM corrected amplitudes and the solid thin curves show the uncorrected amplitudes.
Removing unphysical modulations with our COM correction allows for the
physical amplitude modulations of precessing systems to become more
apparent.  For the COM drift plots on the right, the axes show the coordinate positions of the 
apparent-horizon centers, normalized by the initial total mass of the system $M_0$. The different colored lines
correspond to the Newtonian COM, Eq.~\eqref{COMdef}, at different resolutions. Note that
each resolution for a simulation uses the same initial conditions. The COM values are
plotted for each system from start until a common horizon is found.
}
\end{figure}
}

\newcommand{\figCOMvals}{%
\begin{figure}
  \includegraphics[width = 6.1in]{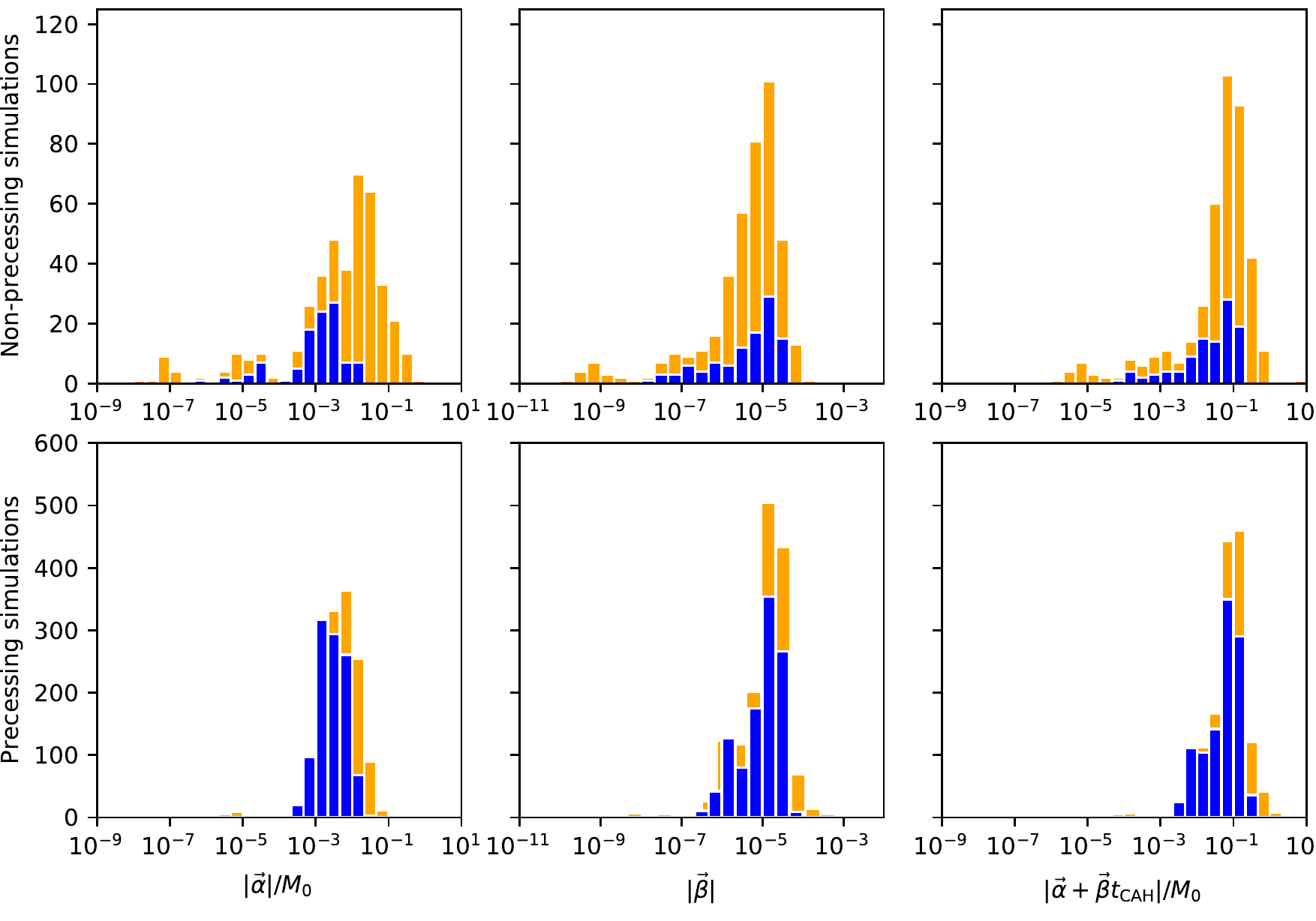}
  \caption{\label{fig:COMvals} Histograms showing the magnitude of the
  center-of-mass (COM) translations $\vec{\alpha}$ and boosts
  $\vec{\beta}$ as defined in Eq.~\eqref{eq:COMsupertranslation_def}, and
total displacements $\vec{\alpha} + t_{\txt{CAH}}\vec{\beta}$, for
all simulations in our catalog.
The top row shows values for non-precessing systems
while the bottom row shows values for precessing systems.
The blue bars denote the newer simulations that utilize the improved initial data procedure~\cite{Ossokine:2015yla}, whereas the
orange bars denote earlier simulations.
}
\end{figure}
}

\newcommand{\figNorbits}{%
\begin{figure}
  \begin{center}
    \includegraphics[width=3.5in]{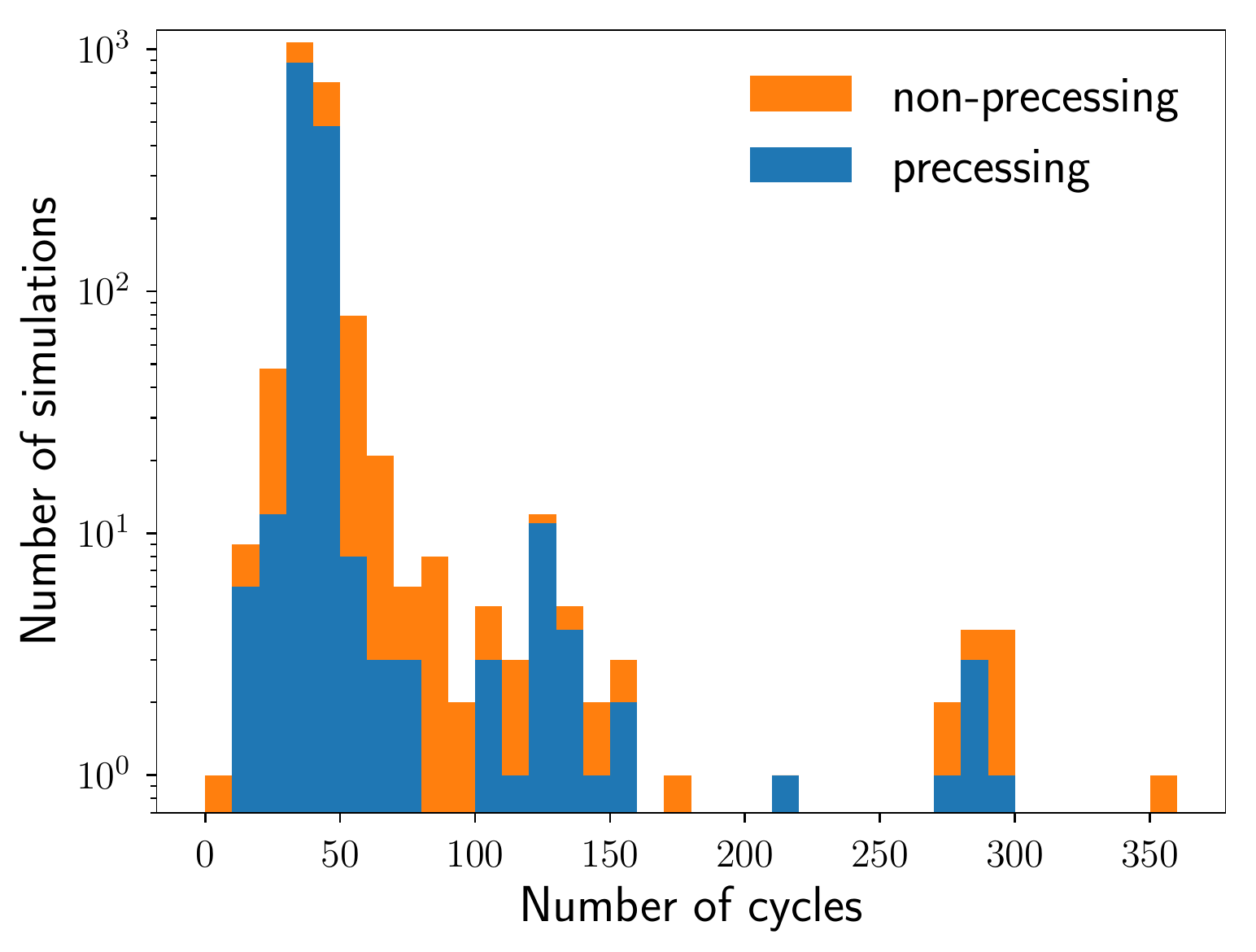}
  \end{center}
  \caption{\label{fig:Norbits}%
    Number of cycles of $\ell=m=2$ gravitational waves before merger
    for the simulations in the catalog, as determined by the
    coordinate trajectories of the black holes.  Bin edges are
    multiples of 10 cycles.
  }
\end{figure}
}

\newcommand{\figEccHist}{%
\begin{figure}
  \begin{center}
    \includegraphics[width=3.5in]{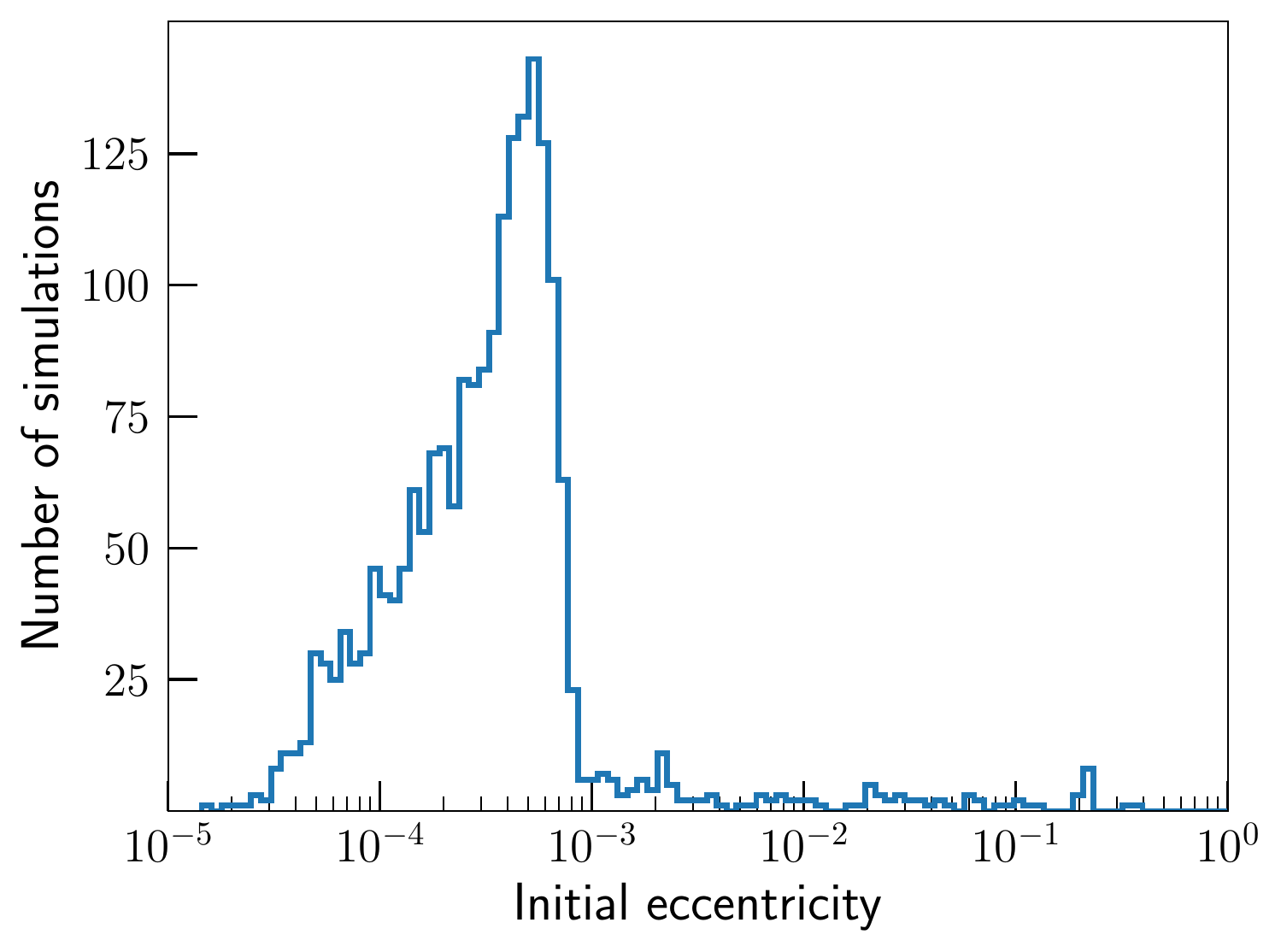}
  \end{center}
  \caption{\label{fig:EccHist}%
    Initial eccentricities $e_{0}$ in the catalog.  The main population
    is the result of eccentricity-reduction, and those
    intentionally exploring high $e_{0}$ constitute the tail.
  }
\end{figure}
}

\newcommand{\figphysicalParamSpace}{%
\begin{figure}
\centerline{  \includegraphics[width=5.5in]{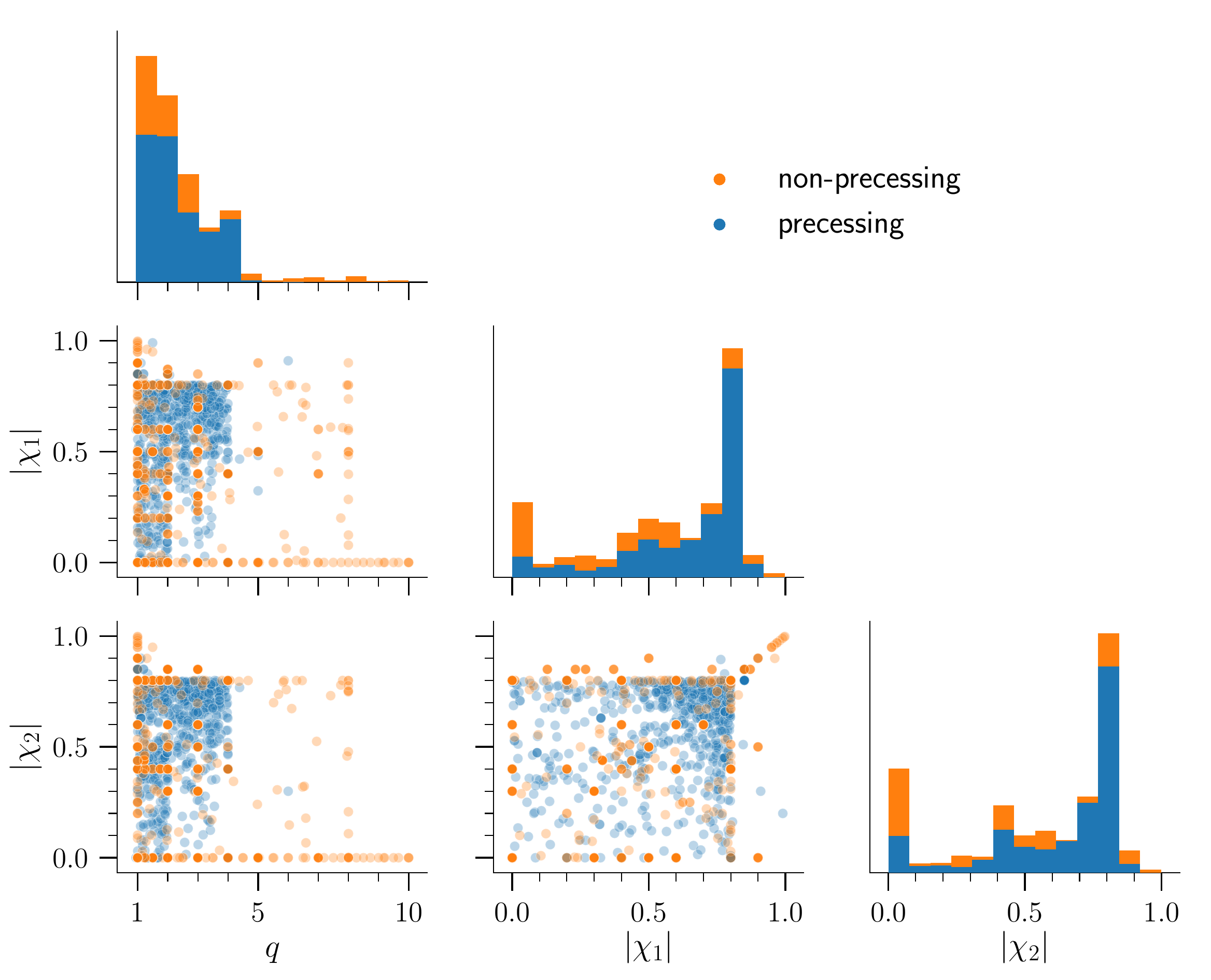}}
  \caption{ \label{fig:physicalParamSpace} Coverage of the SXS Catalog
    parameter space.  Each point is one simulation. Shown here are
    the mass ratio $q = m_{1}/m_{2}$ and the spin magnitudes $|\chi_1|$ and $|\chi_2|$
    of the larger and smaller black hole, respectively.
    Orange points correspond to
    configurations that are not precessing (spins aligned with the
    orbital angular momentum), while blue points correspond to precessing
    configurations.
  }
\end{figure}
}

\newcommand{\figchieffxy}{%
\begin{figure}
  \centerline{\includegraphics[width=5.5in]{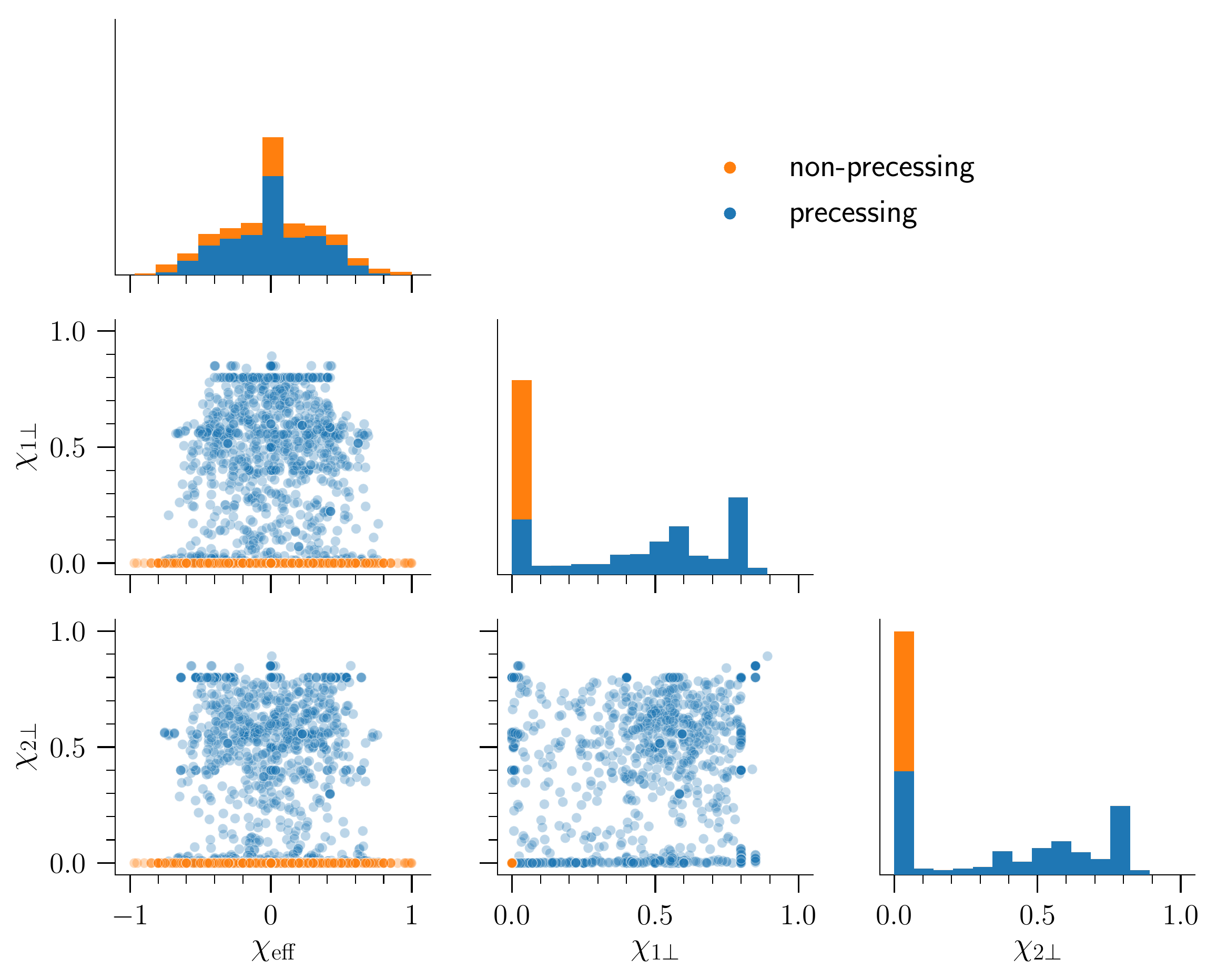}}
  \caption{%
    Distribution of black hole spins
    in the catalog.  Each panel shows a projection of the
    7-dimensional space.  Each point is one simulation. We plot
    the effective spin $\chi_{\rm{eff}}$
    [the combination of spins that has a strong effect on the phasing of the
    gravitational waves; defined in Eq.~\eqref{eq:chieff}] and the magnitudes of the spins in
    the orbital plane.
    Orange points correspond to
    configurations that are not precessing (spins aligned with the
    orbital angular momentum), while blue points correspond to precessing
    configurations. \label{fig:chieffxy}
    }
\end{figure}
}

\newcommand{\figprecAngles}{%
\begin{figure}
  \centerline{\includegraphics[width=5.5in]{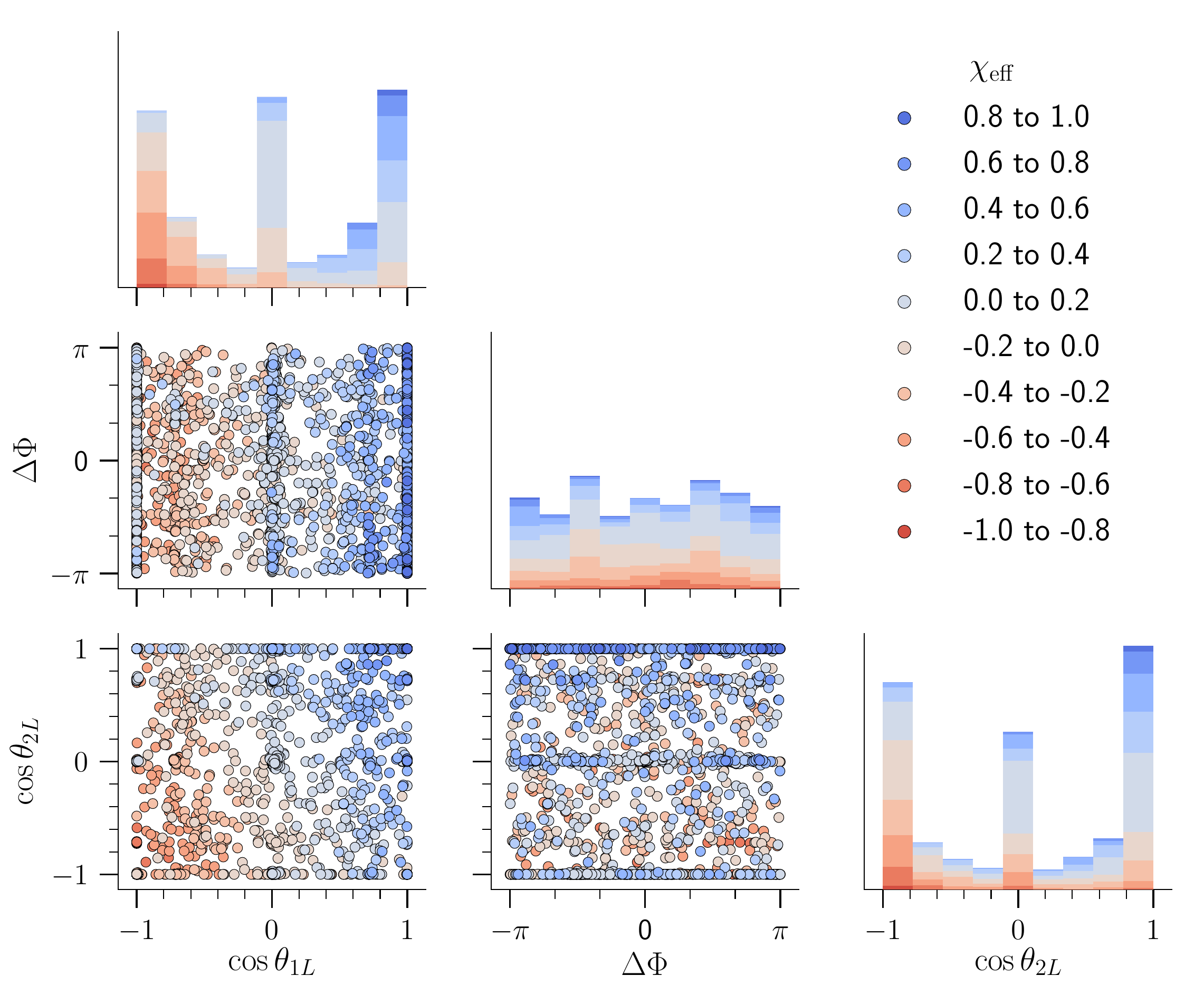}}
  \caption{%
    Projection of parameter space relevant to spin-precession
    dynamics.  Plotted are the cosines of the two tilt angles
    $\theta_{1L}, \theta_{2L}$, and the angle in the orbital plane
    $\Delta\Phi$ between the two spin vectors, all measured at the
    reference time.  Points are colored by the effective spin
    $\chi_{\rm eff}$.
    \label{fig:precAngles}
  }
\end{figure}
}

\newcommand{\figLIGOcomparison}{%
\begin{figure*}
  \begin{center}
    \includegraphics[width=0.49 \columnwidth]{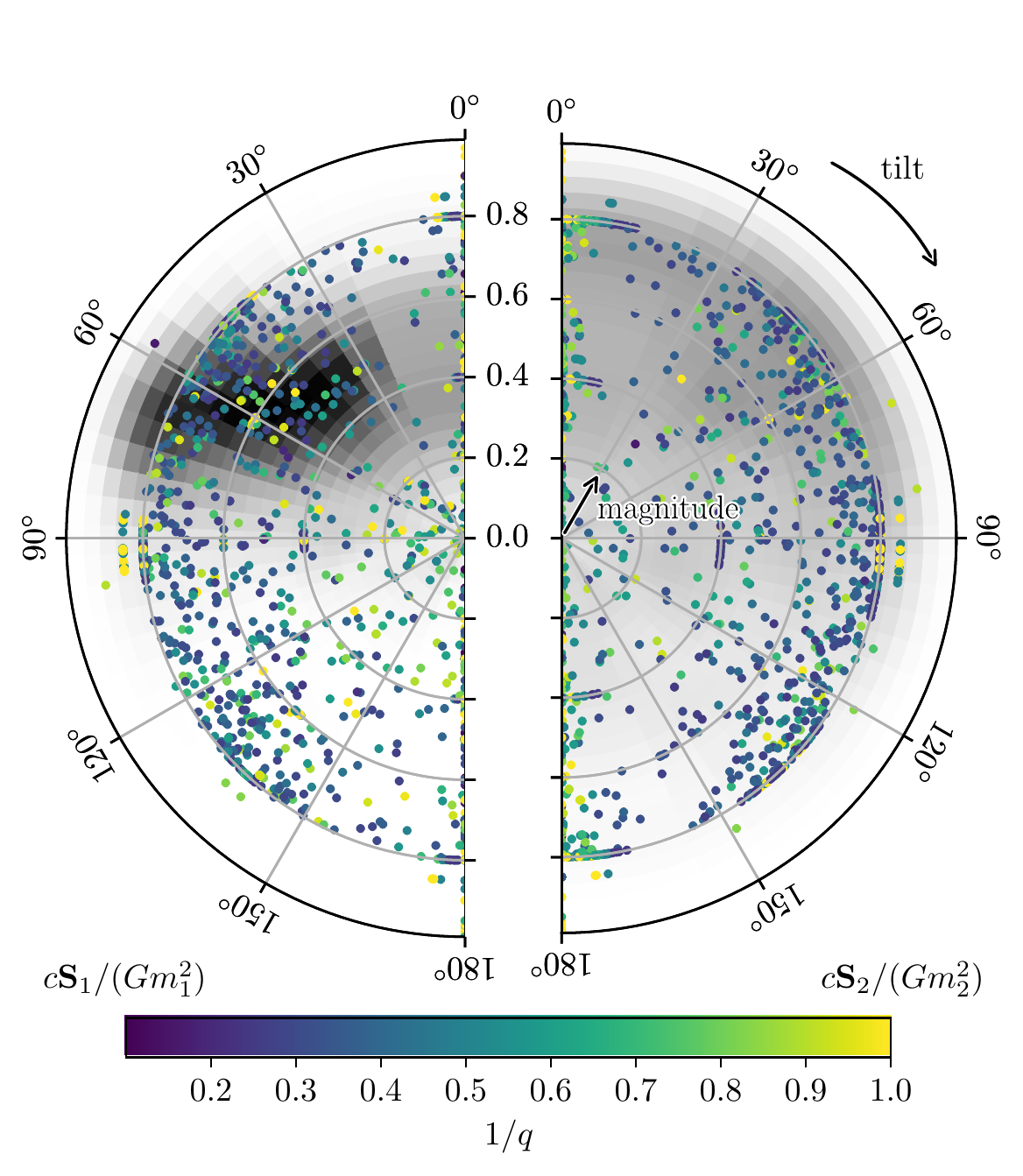}
    \includegraphics[width=0.49 \columnwidth]{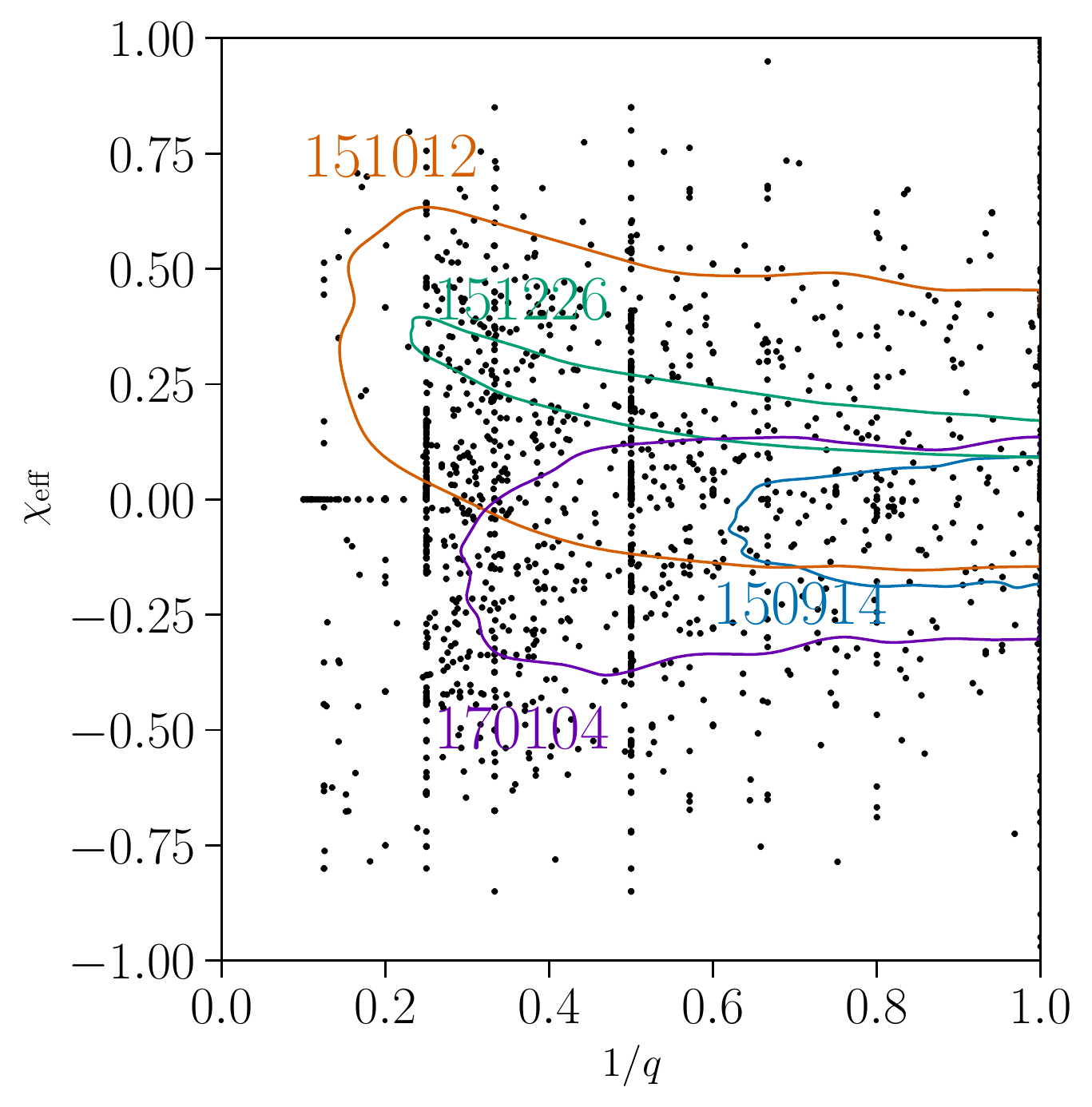}
  \end{center}
  \caption{\label{fig:LIGOcomparison} Parameter space coverage of the SXS catalog compared to
the properties of selected BBH mergers observed through gravitational waves.
(Left) Dimensionless spin for the two binary components. Scatter dots represent the
simulations in the SXS catalog, while the greyscale pixels represent the posterior probability
density as measured for GW151226. (Right) Mass ratio and effective spin. Black dots represent
the simulations in the SXS catalog. The curves are 90\% contours for the 2-dimensional
posterior probability for GW150914 (blue), GW151012 (orange), GW151226 (green), and GW170104
(purple).}
\end{figure*}
}

\newcommand{\figNRMismatch}{%
\begin{figure}
  \includegraphics[width=6in]{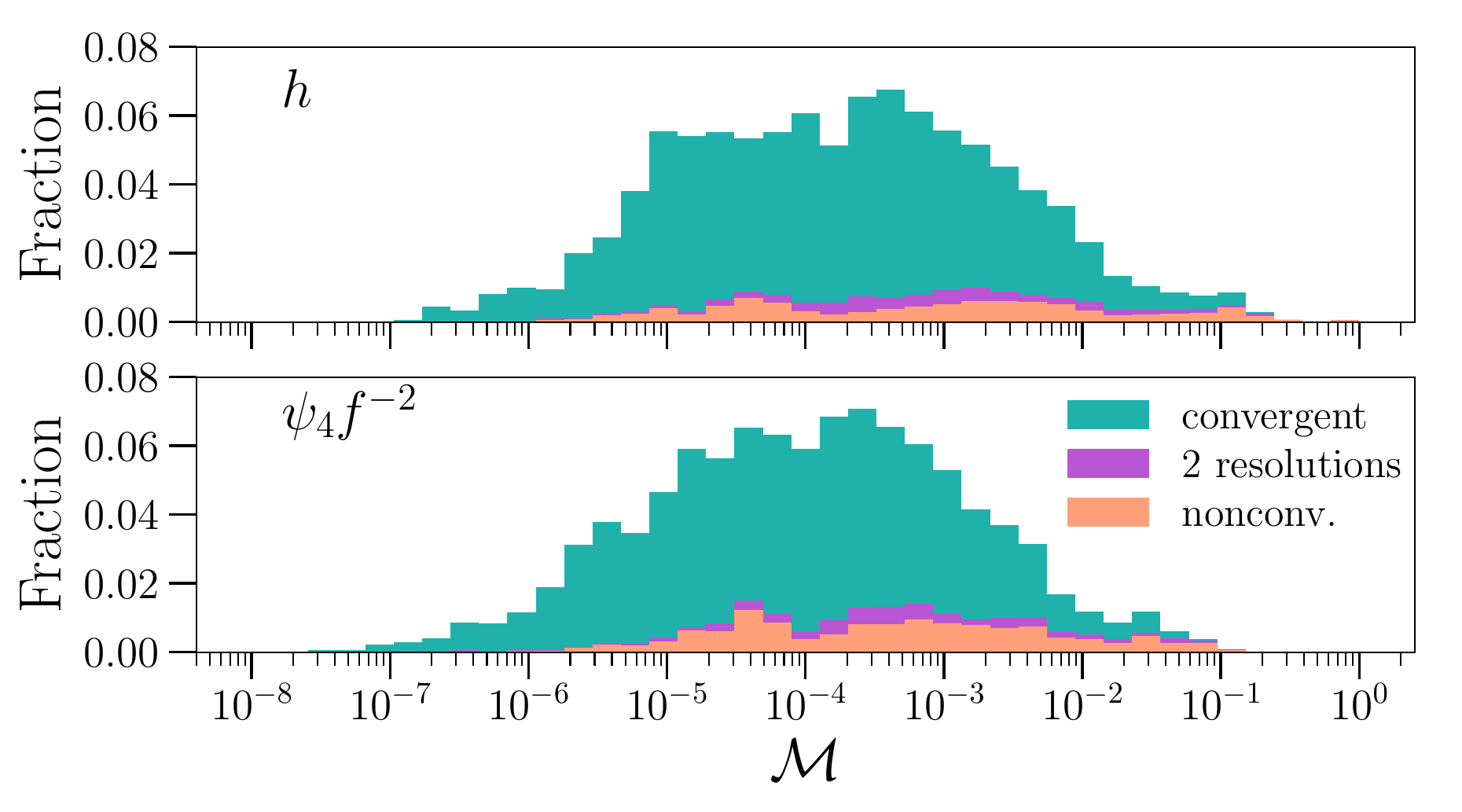}\label{fig:NR_Mismatch}
  \caption{Histogram of flat-noise-curve
    mismatches between the two highest-resolution
    simulations for 20 uniformly distributed detector-direction angles
    for each case in the catalog.
    The horizontal axis represents the mismatch, and the
    vertical axis represents the fraction of all cases with that
    mismatch.  The top plot shows the mismatch between $h$ computed
    via Regge-Wheeler-Zerilli extraction, and the bottom plot shows
    the mismatch between $\Psi_4 f^{-2}$ as computed via Newman-Penrose
    extraction.  The factor of $f^{-2}$ gives the top and bottom plots
    the same frequency weighting.  The
    entries labeled ``convergent'' indicate that the mismatch between
    the two highest resolutions is less than the mismatch between the
    next two highest resolutions; the entries labeled ``nonconv.''
    indicate the opposite.  Cases with only two resolutions are
    so labeled, and cases with only a single resolution are omitted.
    \label{fig:levmismatch}
  }
\end{figure}
}

\newcommand{\figPsihmismatch}{%
\begin{figure}
  \includegraphics[width=6in]{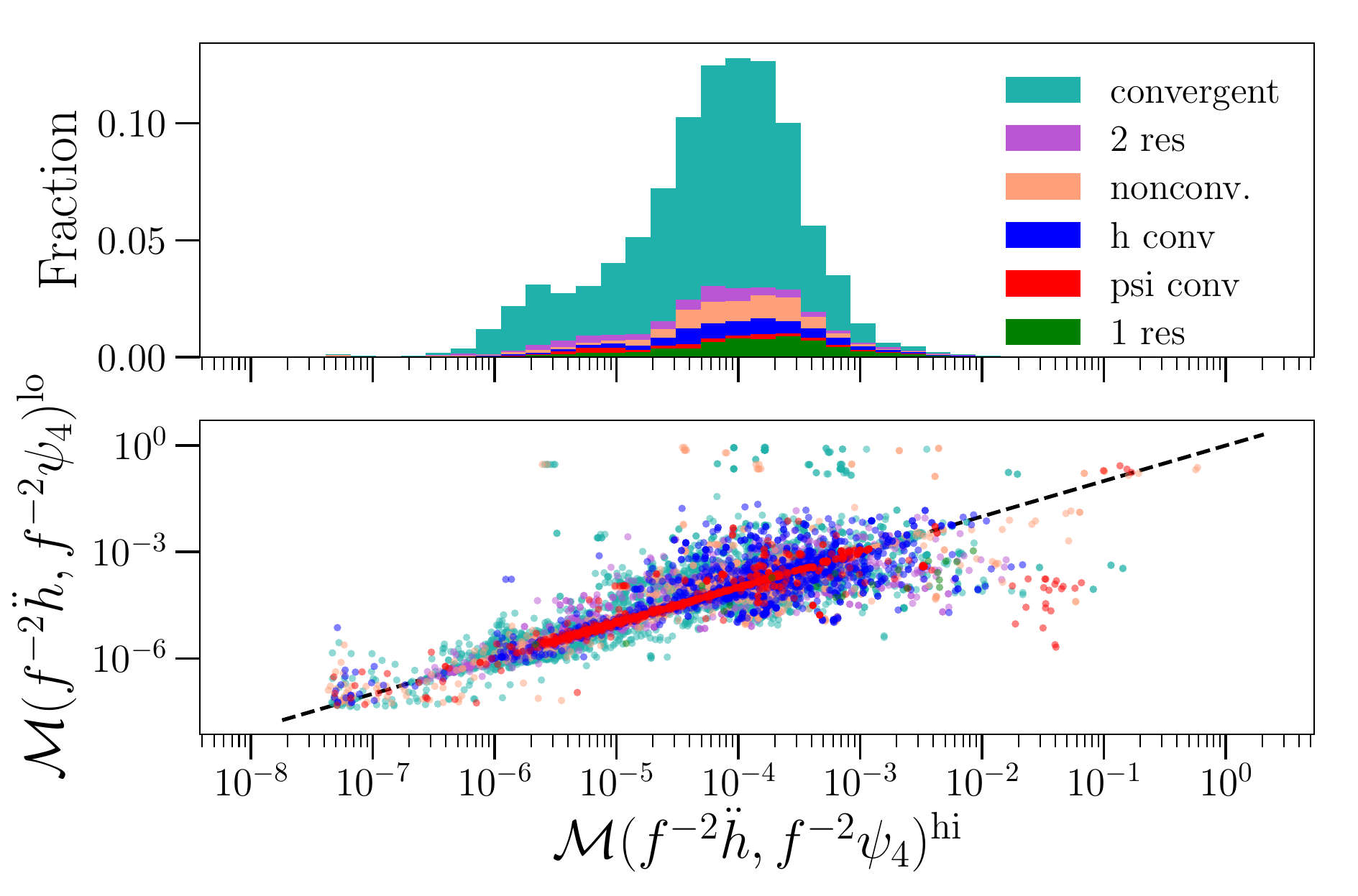}
  \caption{%
    Mismatches, using a flat spectral noise curve, comparing the
    two methods of gravitational wave extraction.
    The top panel depicts histograms of the mismatch between
    $\ddot h f^{-2}$ and $\Psi_4 f^{-2}$ using the highest resolution from each
    simulation. The factor $f^{-2}$ ensures that
    the mismatches here have the same frequency weighting as those of
    Fig.~\ref{fig:levmismatch}.
    The entries are labeled using the same conventions as Fig.~\ref{fig:levmismatch},
    but in addition simulations that are convergent for $h$
    but not $\Psi_4$ and vice versa are labeled, as are simulations with only
    a single resolution.
    The bottom panel depicts a scatter plot of
    the mismatch between $\ddot h f^{-2}$ and $\Psi_4 f^{-2}$ for the highest
    resolution and the same mismatch for the second-highest resolution.
  \label{fig:Psihmismatch}}
\end{figure}
}

\newcommand{\figExtrapOrderMismatch}{%
\begin{figure}
  \includegraphics[width=6in]{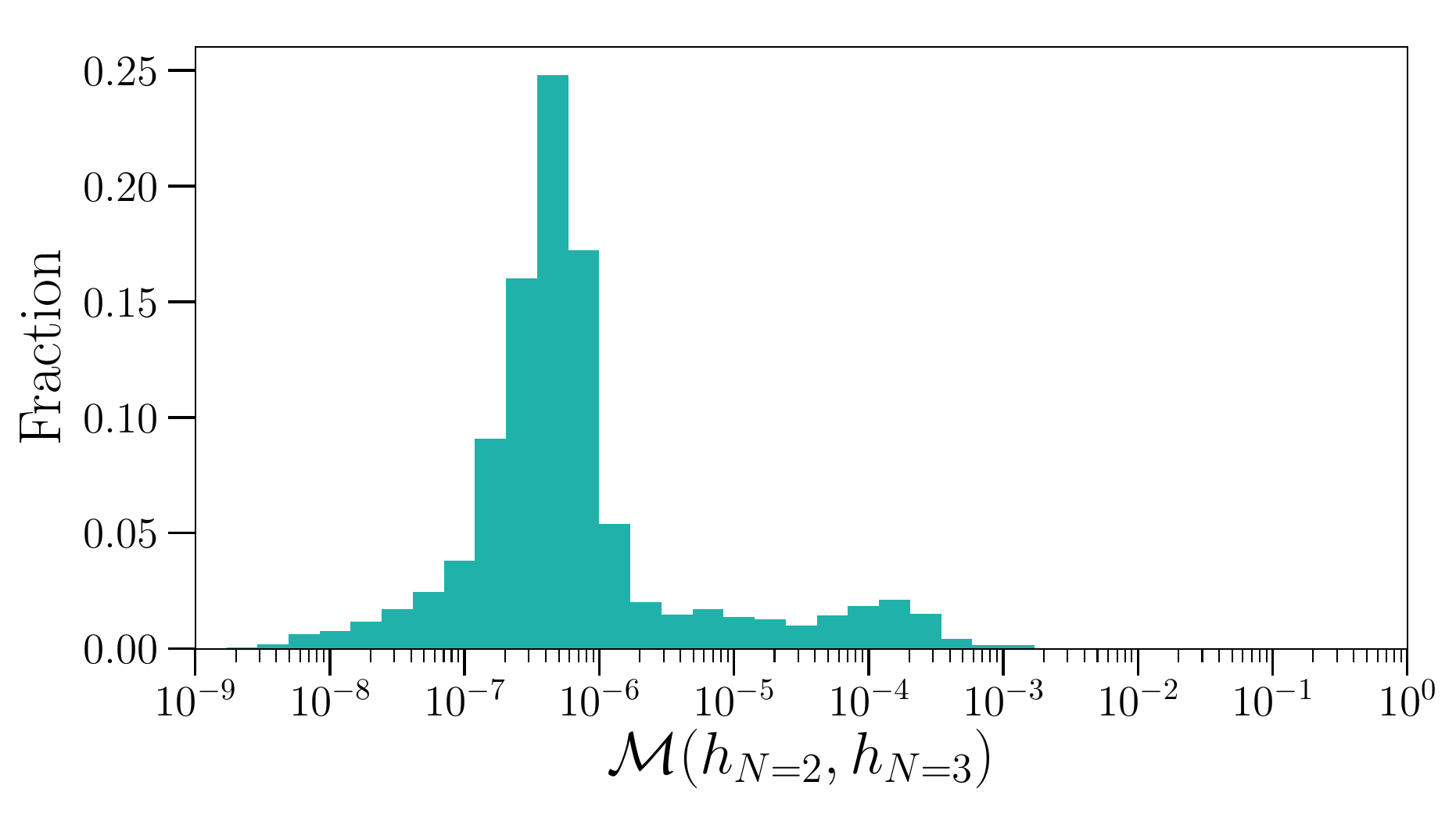} \caption{%
  Mismatches,
  using a flat noise curve, comparing the highest-resolution strain
  waveform of each simulation extrapolated to infinity using
  extrapolation order $N=2$ versus the same waveform extrapolated to
  infinity using order $N=3$.  Thus, the only differences entering the
  mismatches are due to details of the extrapolation procedure.  As in
  Fig.~\ref{fig:levmismatch}, 20 uniformly distributed
  detector-direction angles are compared for each waveform, and all
  mismatches are computed after center-of-mass correction.
  \label{fig:ExtrapOrderMismatch}}
\end{figure}
}

\newcommand{\figMassfits}{%
\begin{figure}
  \begin{center}
  \includegraphics[width=6in]{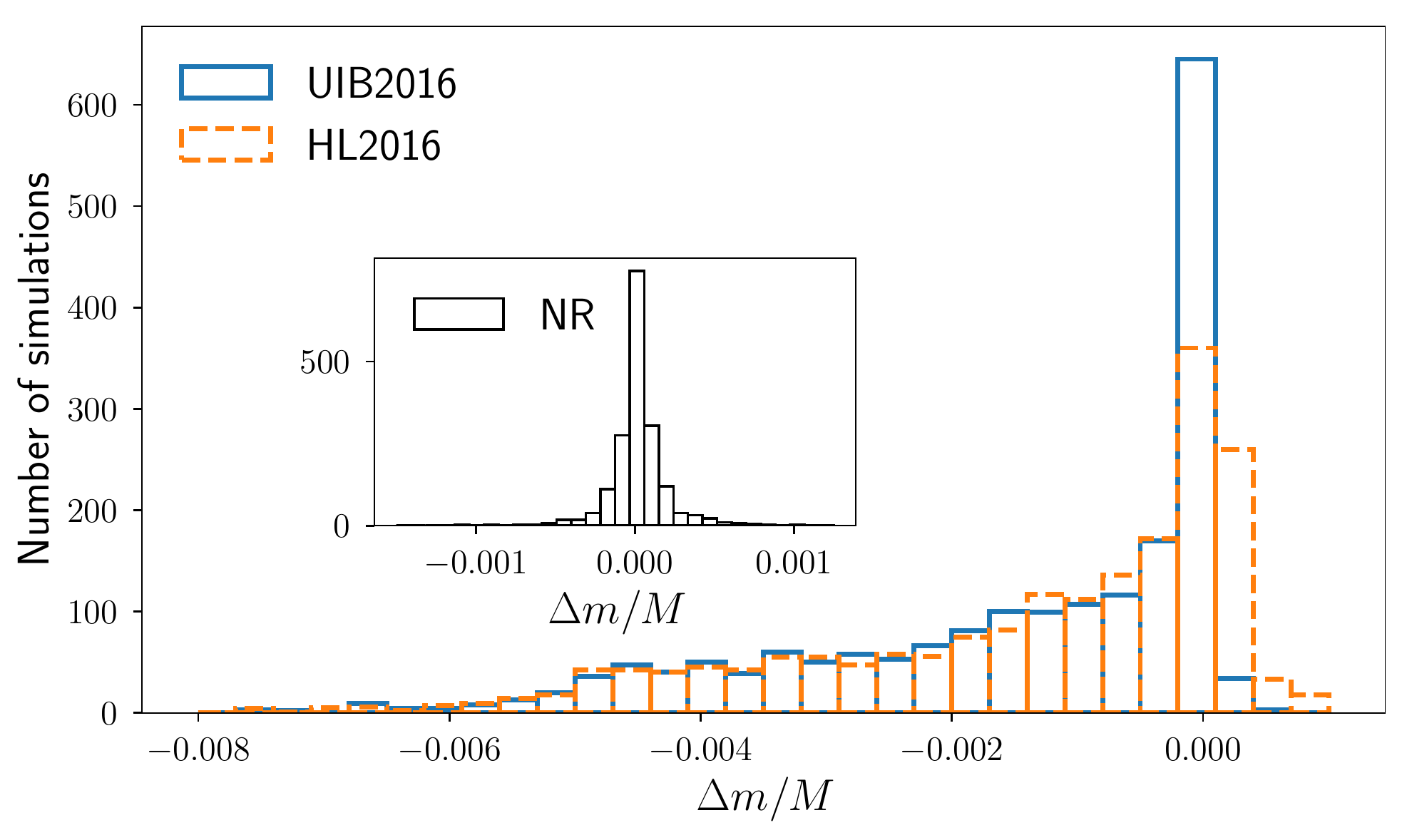}
  \end{center}
  \caption{%
    The difference between NR and two fits for the final
    mass: UIB2016~\cite{Jimenez-Forteza:2016oae} and
    HL2016~\cite{Healy:2016lce}.
    The inset shows our numerical error,
    estimated as the difference
    between the highest and second highest resolutions.
    The accuracy of the fits is excellent, with 90\% of the errors
    $\lesssim{}0.004M$,
    about an order of magnitude larger than our NR errors.%
    \label{fig:Mass_fits}}
\end{figure}
}

\newcommand{\figSpinfits}{%
\begin{figure}
  \begin{center}
  \includegraphics[width=6in]{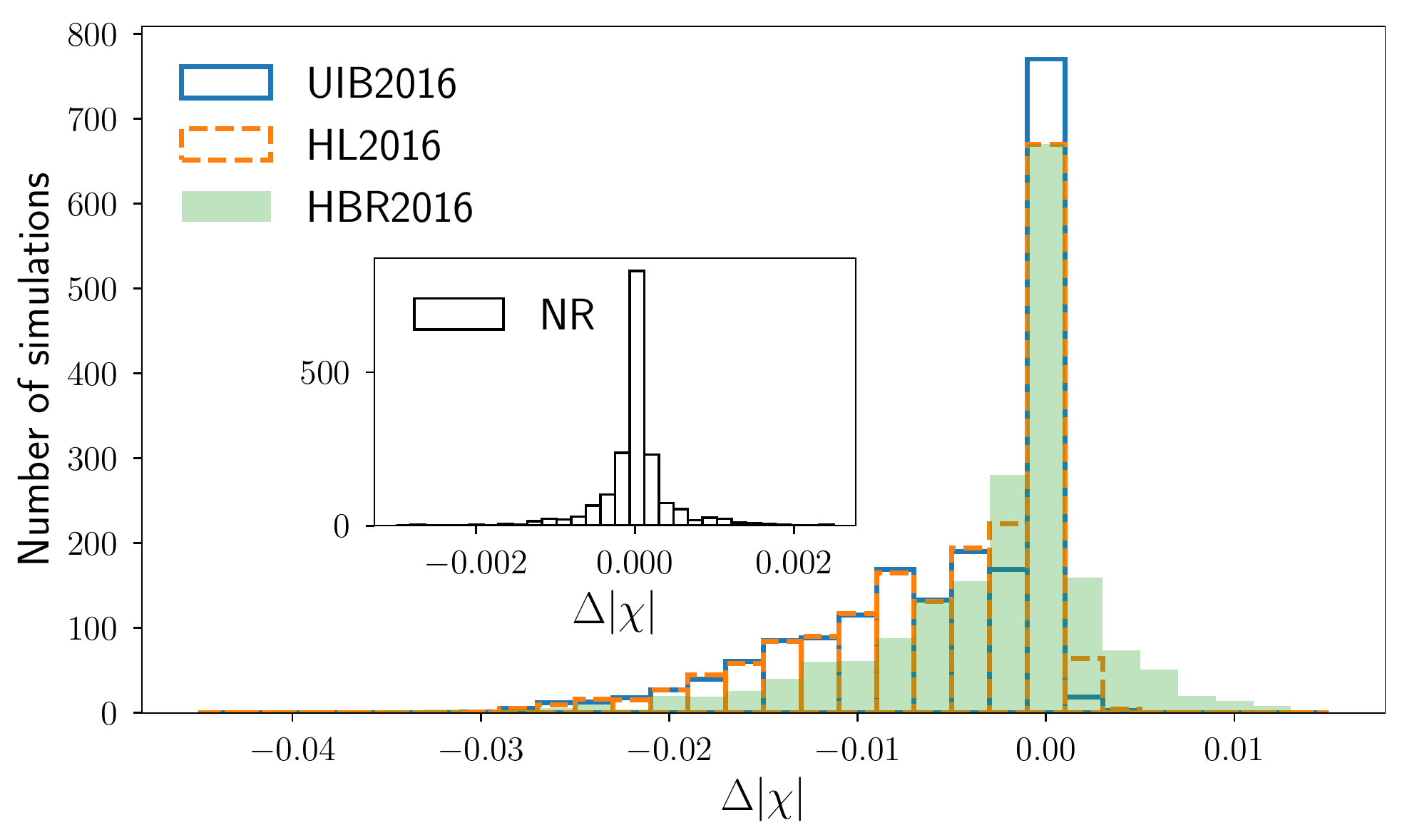}
  \end{center}
  \caption{%
    The difference between NR and three fits for the
    magnitude of the final dimensionless spin:
    UIB2016~\cite{Jimenez-Forteza:2016oae},
    HL2016~\cite{Healy:2016lce}, and HBR2016~\cite{Hofmann:2016yih}.
    The inset shows our numerical
    error. The accuracy of the fits is still good, with 90\% of the errors
    $\lesssim 0.01$,
    but significantly larger
    than our NR errors which are of the order $7\times 10^{-4}$.%
    \label{fig:Spin_fits}}
\end{figure}
}

\newcommand{\macrocolor}[1]{\textcolor{red}{#1}}
\renewcommand{\macrocolor}[1]{#1}
\newcommand{\Nconfigs}{\macrocolor{2018}}
\newcommand{\NconfigsFactor}{\macrocolor{11}}
\newcommand{\Nprec}{\macrocolor{1426}}

\newcommand{\qmin}{\macrocolor{1}}
\newcommand{\qmax}{\macrocolor{10}}
\newcommand{\chimax}{\macrocolor{0.998}}
\newcommand{\medNGWcyc}{\macrocolor{39}}
\newcommand{\minNGWcyc}{\macrocolor{7.0}}
\newcommand{\maxNGWcyc}{\macrocolor{351.3}}

\newcommand{\NRFitError}{\macrocolor{0.03\% and  0.1\%}}

\newcommand{\MADM}{\ensuremath{M_{\textrm{ADM}}}}
\newcommand{\rar}{\ensuremath{r}}
\newcommand{\corotating}[1]{\ensuremath{\hat{#1}}}

\newcommand{\ave}[1]{\ensuremath{\langle#1\rangle}}

\def\yes{\checkmark}
\def\no{\ding{55}}

\newcommand{\txt}[1]{{\textrm{\tiny{#1}}}}

\DeclareMathOperator{\sgn}{sgn}

\begin{document}

\title{The SXS Collaboration catalog of binary black hole simulations}
\newcommand{\AEI}{Albert-Einstein-Institut,
        Max-Planck-Institut für Gravitationsphysik,
    D-14476 Potsdam-Golm, Germany} %
\newcommand{\BHam}{School of Physics and Astronomy and Institute for
  Gravitational Wave Astronomy, University of Birmingham, Edgbaston,
  Birmingham, B15 9TT, UK}
\newcommand{\Caltech}{Theoretical Astrophysics 350-17,
    California Institute of Technology, Pasadena, CA 91125, USA}
\newcommand{\Cornell}{Cornell Center for Astrophysics and Planetary Science,
    Cornell University, Ithaca, New York 14853, USA}
\newcommand{\CITA}{Canadian Institute for Theoretical
    Astrophysics, 60 St.~George Street, University of Toronto,
    Toronto, ON M5S 3H8, Canada} %
\newcommand{\JPL}{Jet Propulsion Laboratory, California Institute of Technology, Pasadena, CA
91109, USA}
\newcommand{\Manchester}{University of Manchester, Manchester, UK}
\newcommand{\MIT}{Department of Physics and MIT Kavli Institute, Cambridge, MA 02139, USA}
\newcommand{\OleMiss}{Department of Physics and Astronomy,
    University of Mississippi, University, MS 38677, USA}
\newcommand{\Radboud}{Department of Astrophysics/IMAPP, Radboud University Nijmegen, P.O. Box
9010, 6500 GL Nijmegen, The Netherlands}
\newcommand{\TorontoPhysics}{Department of Physics
  60 St.~George Street, University of Toronto,
    Toronto, ON M5S 3H8, Canada} %
\newcommand{\Tokyo}{Research Center for the Early Universe, University of Tokyo, Tokyo, 113-0033, Japan}%
\newcommand{\GWPAC}{Gravitational Wave Physics and
    Astronomy Center, California State University Fullerton,
    Fullerton, California 92834, USA} %
\newcommand{\UMass}{Department of Mathematics, University of Massachusetts, Dartmouth, MA
02747, USA} %
\newcommand{\UTA}{Theory Group, Department of Physics, University of Texas at Austin, Austin, TX 78712, USA}

\newcommand{\NsimsMoreThanOneRes}{\macrocolor{1872}}
\newcommand{\NsimsMoreThanTwoRes}{\macrocolor{1777}}
\newcommand{\NsimsTotal}{\macrocolor{1872}}
\newcommand{\NmismatchesTotal}{\macrocolor{37440}}
\newcommand{\NmismatchesNonconvergentH}{\macrocolor{3558}}
\newcommand{\NmismatchesNonconvergentPsi}{\macrocolor{4851}}

\def\aj{\rm{AJ}}
\def\apj{\rm{ApJ}}
\def\apjl{\rm{ApJ}}
\def\apjs{\rm{ApJS}}
\def\aap{\rm{A\&A}}
\def\aaps{\rm{A\&AS}}
\def\mnras{\rm{MNRAS}}
\def\prd{\rm{PRD}}
\def\prl{\rm{PRL}}
\def\prx{\rm{PRX}}
\def\nat{\rm{Nature}}
\def\cqg{\rm{CQG}}
\def\grg{\rm{GRG}}
\def\lrr{\rm{LRR}}

\author{%
Michael~Boyle$^{1}$,
Daniel~Hemberger$^{2}$,
Dante~A.B.~Iozzo$^{1}$,
Geoffrey~Lovelace$^{3,2}$,
Serguei~Ossokine$^{4}$,
Harald~P.~Pfeiffer$^{4}$,
Mark~A.~Scheel$^{2}$,
Leo~C.~Stein$^{5,2}$,
Charles~J.~Woodford$^{6,7}$,
Aaron~B.~Zimmerman$^{8}$,
Nousha~Afshari$^{3}$,
Kevin~Barkett$^{2}$,
Jonathan~Blackman$^{2}$,
Katerina~Chatziioannou$^{7}$,
Tony~Chu$^{7}$,
Nicholas~Demos$^{3}$,
Nils~Deppe$^{1}$,
Scott~E.~Field$^{1,9}$,
Nils~L.~Fischer$^{4}$,
Evan~Foley$^{3}$,
Heather~Fong$^{6,7,10}$,
Alyssa~Garcia$^{3}$,
Matthew~Giesler$^{2}$,
Francois~Hebert$^{2}$,
Ian~Hinder$^{4,11}$,
Reza~Katebi$^{3}$,
Haroon~Khan$^{3}$,
Lawrence~E.~Kidder$^{1}$,
Prayush~Kumar$^{1,7}$,
Kevin~Kuper$^{3}$,
Halston~Lim$^{2,12}$,
Maria~Okounkova$^{2}$,
Teresita~Ramirez$^{3}$,
Samuel~Rodriguez$^{3}$,
Hannes~R.~R{\"u}ter$^{4}$,
Patricia~Schmidt$^{13,2}$,
Bela~Szilagyi$^{2,14}$,
Saul~A.~Teukolsky$^{1,2}$,
Vijay~Varma$^{2}$
and
Marissa~Walker$^{3}$
}

\address{$^{1}$~\Cornell}
\address{$^{2}$~\Caltech}
\address{$^{3}$~\GWPAC}
\address{$^{4}$~\AEI}
\address{$^{5}$~\OleMiss}
\address{$^{6}$~\TorontoPhysics}
\address{$^{7}$~\CITA}
\address{$^{8}$~\UTA}
\address{$^{9}$~\UMass}
\address{$^{10}$~\Tokyo}
\address{$^{11}$~\Manchester}
\address{$^{12}$~\MIT}
\address{$^{13}$~\BHam}
\address{$^{14}$~\JPL}

\clearpage{}

\begin{abstract}
Accurate models of gravitational waves from merging black holes are
necessary for detectors to observe as many
events as possible while extracting the maximum science.
Near the time of merger, the gravitational waves from merging
black holes can be computed only using numerical relativity. In this
paper, we present a major update of the Simulating eXtreme Spacetimes (SXS)
Collaboration catalog of numerical simulations for merging black holes.
The catalog contains \Nconfigs{}
distinct configurations (a factor of \NconfigsFactor{} increase
compared to the 2013 SXS catalog),
including \Nprec{} spin-precessing configurations,
with mass ratios between \qmin{} and \qmax{}, and spin magnitudes up to
\chimax{}. The median length of a waveform in the catalog is
\medNGWcyc{} cycles of the dominant $\ell=m=2$ gravitational-wave mode, with the shortest waveform
containing \minNGWcyc{} cycles and the longest \maxNGWcyc{} cycles.
We discuss improvements such as correcting for moving centers of
mass and extended coverage of the parameter space.
We also present a thorough analysis of numerical errors, finding typical truncation errors corresponding to a waveform mismatch of $\sim 10^{-4}$.
The simulations provide remnant masses and spins with
uncertainties of \NRFitError{} ($90^{\text{th}}$ percentile), about an
order of magnitude better than
analytical models for remnant properties.
The full catalog is publicly available at
\url{https://www.black-holes.org/waveforms}\,.
\end{abstract}

\vspace{-1em}

\section{Introduction}

Advanced LIGO~\cite{TheLIGOScientific:2014jea} and Virgo~\cite{TheVirgo:2014hva} inaugurated the era of gravitational-wave
astronomy in 2015 by observing gravitational waves passing through Earth
for the first time~\cite{Abbott:2016blz, TheLIGOScientific:2016qqj, TheLIGOScientific:2016agk}.
This first gravitational-wave signal, named GW150914, was emitted during the merger
of two black holes~\cite{TheLIGOScientific:2016wfe}.
Subsequently, gravitational-wave signals
have been observed from 
a merger of two neutron stars, GW170817~\cite{TheLIGOScientific:2017qsa}, 
and from nine further black-hole mergers~\cite{Abbott:2016nmj,
  Abbott:2017vtc, Abbott:2017gyy, Abbott:2017oio,
  LIGOScientific:2018mvr}.

Making the most sensitive searches for binary coalescence in noisy
detector data requires accurate gravitational-wave templates.
Further, inferring properties of the sources of these signals requires
comparing the data against millions of accurate templates.
During the late stages of a compact binary merger, when
the components move at relativistic speeds and spacetime becomes nonlinearly dynamical, analytic approximations to the binary dynamics~\cite{Blanchet:2013haa}
break down. This strong-gravity regime reveals the behavior of curved spacetime 
under the most extreme conditions, such as the nonlinear dynamics of
merging black holes, the formation and relaxation of dynamical
horizons~\cite{Ashtekar:2013qta}, and the nature of the remnant black hole left behind
following the merger of a binary black hole (BBH)~\cite{Yang:2017zxs}. The strong-gravity regime also
has the potential to place strong upper bounds on
deviations from general relativity (or to reveal such deviations if they exist)~\cite{Berti:2015itd,TheLIGOScientific:2016src,
Berti:2018vdi,Berti:2018cxi}.

In this highly nonlinear regime, accurate solutions of Einstein's
equations of general relativity require numerical-relativity
calculations: direct solution of the full dynamical field equations
using high-performance computing (for summaries,
see~\cite{Baumgarte:2010ndz, Lehner:2014asa, Cardoso:2014uka, Duez:2018jaf} and references therein),
which became possible in 2005~\cite{Pretorius:2005gq,
  Campanelli:2005dd, Baker:2005vv}.
Such simulations are essential in exploring the dynamics of spacetime
curvature itself. They have revealed the simplicity of the merger
phase~\cite{Buonanno:2006ui} and the potentially strong recoil of the
remnant (e.g.~\cite{Campanelli:2007cga,Gonzalez:2007hi,Lousto:2011kp}),
motivating studies of the interplay between the linear momenta of the
black holes and of the surrounding
spacetime~\cite{Keppel:2009tc,Lovelace:2009dg}.
Simulations have also been used for visualizations of curved spacetime~\cite{Owen:2010fa,Zimmerman:2011as,Nichols:2011pu,Dennison:2012ue,Dennison:2012vf,Zhang:2012jj,Nichols:2012jn,Price:2012uh},
investigations of spin quantities~\cite{Lovelace:2014twa}, and the relaxation of spacetime to the Kerr solution following merger~\cite{Owen:2009sb,Owen:2010vw,Bhagwat:2017tkm}. 
The motion of the black hole horizons and horizon curvature quantities
have been used to explore eccentric dynamics~\cite{Mroue:2010re,Buonanno:2010yk,Mroue:2012kv,Lewis:2016lgx}, spin precession~\cite{Ossokine:2015vda,Lousto:2015uwa,Owen:2017yaj,Lousto:2018dgd}, and the first law of binary black hole mechanics~\cite{LeTiec:2011ab,Tiec:2015cxa,Blanchet:2012at,Zimmerman:2016ajr,LeTiec:2017ebm}.
These in turn have been compared to analytic post-Newtonian and self-force approximations (see also~\cite{LeTiec:2011bk,Tiec:2013twa,vandeMeent:2016hel}), mapping out the bounds of validity of these approximations.

A key application of BBH simulations is the accurate modeling of
gravitational waves emitted by these systems during their late
inspiral, merger, and final ringdown.  Waveforms extracted from BBH
simulations are essential for analyzing observed gravitational-wave
signals from black hole binaries.  Indeed, all BBH observations by
LIGO and Virgo were analyzed using waveform families that rely on
numerical relativity for their construction, most notably
effective-one-body waveform
models~\cite{Buonanno:1998gg, Buonanno:2000ef, Taracchini:2013rva,
  Bohe:2016gbl, Pan:2013rra}
and phenomenological waveform
models~\cite{Hannam:2013oca,Khan:2015jqa,Husa:2015iqa}.  Numerical
simulations are also central in validating such waveform
models~\cite{Damour:2007yf, Boyle:2008ge, Pan:2013tva, Kumar:2015tha,
  Kumar:2016dhh, Babak:2016tgq, Abbott:2016wiq, Lovelace:2016uwp}, and
were used to validate GW
searches~\cite{Aylott:2009ya,Aasi:2014tra,Abbott:2016apu}.  Waveforms
from numerical relativity are also used directly in parameter
estimation~\cite{Heal:2017abq, Lange:2017wki}, to construct template
banks~\cite{Kumar:2013gwa}, and to construct waveform families without
intermediate analytical models, using methods such as reduced order
modeling~\cite{Blackman:2017pcm, Varma:2018mmi, Blackman:2017dfb,
  Blackman:2015pia}.
Today's simulations span ground-based detectors' frequency bands only
for total masses $\gtrsim 50 M_\odot$. For smaller total masses,
numerical-relativity waveforms can be ``hybridized'' by attaching them
to the end of analytic inspiral waveforms to produce waveforms that
span detectors' sensitive bands~\cite{Varma:2018mmi, Boyle:2011dy,
  MacDonald:2011ne, Ohme:2011zm, Hannam:2010ky}.

Most applications of numerical relativity to 
gravitational-wave astronomy~\cite{Tichy:2008du,
Blackman:2015pia,Bohe:2016gbl,Khan:2015jqa,Abbott:2016apu}
require merger simulations for a large number of different 
BBH configurations. Several groups have answered this challenge by creating public catalogs of numerical-relativity simulations for merging black holes~\cite{SXSCatalog, RITCatalog, GaTechCatalog}.
Even with this considerable progress, computing simulation
catalogs that meet the needs of current and future gravitational-wave
observatories remains quite challenging. Simulations must be long
enough to span LIGO's sensitive frequency band or to allow
  reliable hybridization.  This is
straightforward for black-hole binaries with sufficiently high total
masses (such as GW150914, which remained in LIGO's band for only $\sim$0.2
seconds~\cite{Abbott:2016blz}) but not yet possible for BBHs
with lower masses (such as GW170608, which remained in LIGO's
band for $\sim$2 seconds~\cite{Abbott:2017gyy}). Moreover, simulations must achieve
sufficient accuracy [as discussed further in the text surrounding
  Eq.~\eqref{eq:mismatchSNR}], which is especially challenging
when computing the waves' higher-order spherical-harmonic modes.
Additionally, catalogs must span a vast, 7-dimensional
parameter space, including the different possible mass ratios and
black-hole spins. While a small number of simulations with extremely high
mass ratios (up to $m_1/m_2 \sim 100$~\cite{Lousto:2010ut,Sperhake:2011ik}) and nearly extremal
spins~\cite{Scheel:2014ina}
have been achieved, much of the high-mass-ratio, high-spin
parameter space remains completely unexplored.

Early efforts to explore the extensive parameter space
of BBHs  with numerical relativity highlighted the importance of collaborative efforts for
development of numerical-relativity codes, production of numerical-relativity waveforms, error estimation,
analysis of the resulting waveforms, and in storing and making broadly available the resulting data.
This in turn inspired the creation of several collaborative projects rooted in numerical relativity and
gravitational wave analysis, with the goal of creating the simulation catalogs that
gravitational-wave astronomy requires. The
first such collaboration resulting in a catalog was
the Numerical INJection
Analysis (NINJA) project that began in 2008~\cite{Aylott:2009tn}. The goal of NINJA was
to test how well ground-based gravitational-wave detectors could find gravitational waves in their data,
by injecting realistic gravitational-wave signals (created from numerical-relativity waveforms) into
detector noise~\cite{Aylott:2009ya, Aylott:2009tn}. NINJA was a pioneering example of collaboration
between members of the numerical relativity and gravitational wave analysis communities. Its catalog
included simulations generated with a variety of numerical-relativity codes,
including BAM~\cite{Bruegmann:2006at}, CCATIE~\cite{Koppitz:2007ev,Schnittman:2007ij,Rezzolla:2007xa,
Alcubierre:2000xu,Alcubierre:2002kk}, Hahndol~\cite{Imbiriba:2004tp,vanMeter:2006vi},
LAZEV~\cite{Zlochower:2005bj}, LEAN~\cite{Sperhake:2006cy}, MayaKranc~\cite{Herrmann:2006ks,Pekowsky:2013ska},
the Princeton University code~\cite{Pretorius:2004jg,Pretorius:2005gq,Buonanno:2006ui,Pretorius:2007jn},
the University of Illinois code~\cite{Etienne:2007hr}, and
our code SpEC~\cite{SpECwebsite} (cf.~Section~\ref{sec:SpECsum}).
Members of the NINJA project created a catalog of 23 simulations for
creating injected signals, but this initial catalog spanned a 
very limited subspace of the 7-dimensional BBH parameter space.

In 2012, a followup effort, ``NINJA-2'' yielded a catalog~\cite{Ajith:2012az, Aasi:2014tra} of 63 simulations spanning a broader region
of the parameter space, but still not including any simulations with precessing black-hole spins.
As the simulations were created by various groups and collaborations, the simulations in the NINJA-2
catalog use a variety of codes, including SpEC~\cite{SpECwebsite} as well as the moving puncture/BSSN~\cite{Baumgarte:1998te}
codes
BAM~\cite{Bruegmann:2006at}, LAZEV~\cite{Zlochower:2005bj}, LEAN~\cite{Sperhake:2006cy},
Llama~\cite{Pollney:2009yz}, and Maya/MayaKranc~\cite{Herrmann:2006ks,Pekowsky:2013ska}.
Note that many of the moving puncture codes utilize a common Cactus
infrastructure~\cite{Cactus} and the Einstein
Toolkit~\cite{Loffler:2011ay}, a finite-volume discretization code.
In 2013 the Numerical Relativity Analytical
Relativity (NRAR) collaboration~\cite{Hinder:2013oqa} computed and published 25 simulations, with the main focus being the
comparison of the numerical relativity waveforms with the analytical waveform families in use by LIGO.

Building on the NINJA and NRAR efforts, a number of numerical-relativity groups have
begun building larger, more comprehensive catalogs, spanning more of the
parameter space (including spin precession) and including more orbits before merger
(enabling the waveforms to span the detectors' frequency bands for lower total masses).
These catalogs are summarized in Table~\ref{table:Catalog_Comp}.
In 2013 the SXS collaboration released a catalog~\cite{Mroue:2013xna}
with 174 BBH simulations created using
the Spectral Einstein Code (SpEC)~\cite{SpECwebsite}.
By the end of 2018, this catalog had grown to include 337
simulations for BBHs, seven from binaries with a black hole and a
neutron star, and two from a pair of neutron stars.
The full SXS catalog (including the new simulations presented here) is
publicly accessible at~\url{https://www.black-holes.org/waveforms}~\cite{SXSCatalog}.
In May 2016 the Georgia Tech group released a catalog~\cite{GaTechCatalog}
of 452 distinct BBH simulations
(from a pool of more than 600 BBH simulations~\cite{Jani:2016wkt}).
The Rochester Institute of Technology (RIT) group released a catalog~\cite{RITCatalog}
in 2017 that included 126 simulations~\cite{Healy:2017psd}, as
well as an updated catalog in 2019 containing a total of 320
simulations~\cite{Healy:2019jyf}.

\begin{table}
\begin{center}

\newcommand{\tabrotang}{80}
\newcommand{\tabcolhead}[1]{\multicolumn{1}{c}{\rotatebox{\tabrotang}{#1}}}
  \begin{tabular}{r|ccrlllcrc}
    \tabcolhead{Catalog} &
    \tabcolhead{Started} &
    \tabcolhead{Updating?} &
    \tabcolhead{Simulations} &
    \tabcolhead{$m_1/m_2$ range} &
    \tabcolhead{$\vert\chi_1\vert$ range} &
    \tabcolhead{$\vert\chi_2\vert$ range} &
    \tabcolhead{Precessing?} &
    \tabcolhead{${\rm Median\, } N_{\rm cyc}$} &
    \tabcolhead{Public?} \\
    \noalign{\smallskip} \hline \noalign{\smallskip} \hline \noalign{\smallskip}
NINJA~\cite{Aylott:2009tn,Ajith:2012az}     &       2008 & \no       &        63 &            1--10  &                  0--0.95 &                  0--0.95 & \no         &           15 & \no    \\
NRAR~\cite{Hinder:2013oqa}       &       2013 & \no       &   25  &            1--10  &                   0--0.8  &                   0--0.6 & \yes        &     24  & \no    \\
Georgia Tech~\cite{Jani:2016wkt}     &       2016 & \yes      &       452 &           1--15  &                   0--0.8 &                   0--0.8 & \yes        &            4  & \yes   \\
RIT (2017)~\cite{Healy:2017psd} &       2017 & \yes      &       126 &           1--6  &                  0--0.85 &                  0--0.85 & \yes        &           16  & \yes   \\
RIT (2019)~\cite{Healy:2019jyf} &       2017 & \yes      &       320 &           1--6  &                  0--0.95  &                  0--0.95  & \yes        &           19  & \yes   \\
NCSA (2019)~\cite{Huerta:2019oxn} &       2019 & \no      &       89 &           1--10 &                 0 &                 0 & \no        &           20 & \no   \\
SXS (2018) &       2013 & \yes      &       337 &           1--10 &                 0--0.995 &                 0--0.995 & \yes        &           23 & \yes   \\
SXS (2019) &       2013 & \yes      & \Nconfigs &    \qmin--\qmax &               0--\chimax &               0--\chimax & \yes        &   \medNGWcyc & \yes   \\
    \noalign{\smallskip} \hline\noalign{\smallskip} \hline\noalign{\smallskip}
  \end{tabular}

\caption{A comparison of BBH simulation catalogs. The mass of the larger black hole is $m_1$, and the mass of the smaller black hole is $m_2$. We use the convention that the mass ratio is $m_1/m_2 >1$.
The dimensionless spin magnitudes of the black holes are denoted
$|\chi_{1,2}|$.
The ``SXS (2018)'' row corresponds to the number of publicly available
simulations at the end of 2018.
\label{table:Catalog_Comp}
}

\end{center}
\end{table}

In this paper, we present a major update to the SXS Collaboration's
catalog. Our catalog, created using SpEC,
now includes \Nconfigs{}
simulations, an increase of a factor of \NconfigsFactor{} over our 2013 catalog.
The median waveform length is now $N_{\rm cyc} =$ \medNGWcyc{} cycles
of $\ell=m=2$ gravitational waves, while in our initial catalog~\cite{Mroue:2013xna},
only half of the simulations had more than 24
gravitational-wave cycles. Here $N_{\rm cyc}$ is approximated by doubling the number
of orbits during inspiral up to merger (when a common horizon forms), as determined
by the coordinate trajectories of the black holes.
The increased number of cycles means that typical waveforms in our catalog
now tend to span LIGO's sensitive band over a broader
range of total masses.  We estimate the numerical
uncertainty for most waveforms in the catalog, finding a typical
mismatch of $\sim 10^{-4}$ between the highest and second-highest
resolutions.
Our catalog now includes spins up to \chimax{} and mass ratios up to \qmax{}, with
considerably better coverage in the space of mass ratios up to 2 and spins up
to 0.8 (motivated by the estimated properties of GW150914). 
In addition we have re-run a number of earlier simulations
with a more modern version of SpEC. In some cases this improves
the precision of higher order modes and removes the imprint of gauge changes from 
the coordinate trajectories.
Since our past simulations have been widely used, we retain all versions of the simulations 
in our catalog with different labels; further details are given in App.~\ref{sec:sxs-catalog-contents}.
Our catalog is publicly available at
\url{https://www.black-holes.org/waveforms}~\cite{SXSCatalog}, in a
format based on that of the NRAR project~\cite{Brown2007}.

The rest of this paper is organized as
follows. Section~\ref{sec:methods} summarizes the numerical methods
that we employ in SpEC. Then, Sec.~\ref{sec:paramSpace} summarizes
the areas of the parameter space our catalog covers (and what areas
it does not yet cover). Section~\ref{sec:waveQuality} estimates the
accuracy of the catalog's waveforms. After comparing the remnant
properties to analytic fits in Sec.~\ref{sec:remnant}, we conclude
in Sec.~\ref{sec:conclusion}.
We document the formats of our publicly available data in
App.~\ref{sec:sxs-catalog-contents}, our definitions for calculation
of mismatches in App.~\ref{app:mismatches}, and sign conventions in App.~\ref{app:signs}.

\section{Summary of methods}\label{sec:methods}

\subsection{Spectral Einstein Code}
\label{sec:SpECsum}

We use the Spectral Einstein Code (SpEC)~\cite{SpECwebsite} to model
merging black holes and the gravitational waves they emit.
The first step in a binary black-hole simulation is constructing
initial data.
We construct constraint-satisfying
BBH initial data using the Extended Conformal Thin
Sandwich (XCTS)~\cite{York:1998hy, Pfeiffer:2002iy} equations. In the
XCTS formulation of the BBH initial value problem,
the initial spatial slice has i) a spatial metric proportional to a freely chosen
conformal metric and ii) a freely chosen trace of its extrinsic curvature. Typically,
when solving the XCTS equations, we choose the conformal metric and trace of extrinsic curvature to be weighted superpositions of the
analytic solutions for two single black holes in Kerr-Schild
coordinates~\cite{Lovelace:2008tw}, but some (typically older) simulations in the SXS
catalog instead
are conformally flat and have a vanishing trace of the extrinsic curvature (i.e.~maximal
slicing)~\cite{Caudill:2006hw}. In the XCTS formulation, the conformal metric and
trace of the extrinsic curvature have freely chosen time derivatives; we construct
quasi-equilibrium initial data by setting these time derivatives to zero.
We then solve the XCTS equations on a grid with two excised
regions using a spectral elliptic
solver~\cite{Pfeiffer:2002wt}, with boundary conditions on the excision
boundaries chosen to ensure that these boundaries are apparent
horizons~\cite{Caudill:2006hw, Lovelace:2008tw}. The solution yields initial data for
a BBH evolution, including the initial spatial metric,
the initial extrinsic curvature, and the initial lapse and shift, which determine the
initial coordinate choice.

We iteratively construct BBH initial data, tuning the initial data to
achieve a BBH with the desired properties.  Our iterative scheme uses two nested loops. The inner loop solves the XCTS equations, adjusting our choices
for the free data (conformal metric, trace of
extrinsic curvature, and the time derivatives of each) and boundary
conditions, until the resulting BBHs have the desired mass ratio
and spins~\cite{Buchman:2012dw,Ossokine:2015yla}. The outer loop briefly (typically for a few orbits)
evolves the initial data resulting from the inner loop,
 and adjusts the
initial coordinate velocities to yield a BBH with small orbital
eccentricity~\cite{Pfeiffer:2007yz, Buonanno:2010yk, Mroue:2012kv},
typically $e_{0} \sim 10^{-4}$ as defined in Eq.~\eqref{eq:e}.
For some simulations in the SXS catalog, we intentionally omit the eccentricity-reduction
loop, to obtain initial data for BBHs with non-negligible orbital eccentricity.

We evolve the initial data using a first-order
version of the generalized harmonic formulation~\cite{Friedrich1985,
  Garfinkle:2001ni, Pretorius:2004jg, Lindblom:2005qh} of Einstein's equations
with constraint
damping~\cite{Gundlach:2005eh, Pretorius:2004jg, Lindblom:2005qh}.  We choose an
initial gauge that approximates a time-independent
solution in a co-rotating frame, and then we smoothly change to damped harmonic
gauge~\cite{Lindblom:2009tu, Choptuik:2009ww, Szilagyi:2009qz}, which we have found
to work well numerically near the time of merger.

We evolve the initial data using a multidomain spectral
method~\cite{Kidder:1999fv, Lindblom:2005qh, Scheel:2008rj,
  Szilagyi:2009qz, Hemberger:2012jz}. Timestepping
is done via the method of lines using
a fifth-order Dormand-Prince integrator
with a proportional-integral adaptive
timestepping control system that chooses an appropriate step size while
achieving a desired time-stepping error~\cite{Press:2007zz}.
The computational domain extends from pure-outflow
excision boundaries conforming to the shapes of the apparent
horizons (AH)~\cite{Scheel:2008rj, Szilagyi:2009qz, Hemberger:2012jz, Ossokine:2013zga}
to an artificial outer boundary where we impose constraint-preserving
boundary conditions~\cite{Lindblom:2005qh, Rinne:2006vv, Rinne:2007ui}. We
also impose boundary conditions on the incoming characteristic fields at
each internal boundary of the computational domain~\cite{HESTHAVEN200023, Bjorhus1995}.
After a common AH forms, the simulation automatically
stops, interpolates onto a new grid with only a single excision
boundary~\cite{Scheel:2008rj, Hemberger:2012jz}, and continues evolving
on this new domain through ringdown, until the ringdown gravitational waves have left the
domain.

Spectral methods are exponentially convergent, i.e.\ spatial
truncation errors in a given subdomain of fixed size and shape
decrease exponentially with the number of collocation points.
Our simulations use
multiple subdomains, and the size, shape, and even the number of these
subdomains changes dynamically as a simulation proceeds ($h$-refinement), as controlled
by our spectral adaptive mesh refinement (AMR)
procedure~\cite{Lovelace:2010ne, Szilagyi:2014fna}.  Moreover, we choose the accuracy
of a simulation not by picking the number of grid points directly, but
instead by specifying a tolerance parameter that governs when AMR
should add or subtract grid points within a given subdomain
($p$-refinement), or when it should
split or join subdomains ($h$-refinement).
As a result, we should not
always expect strict convergence as a function of the AMR
tolerance parameter.  Convergence may fail in several ways.
For example, two otherwise-identical simulations with different
AMR tolerances may happen to have the same number of grid
points in a particular subdomain at a particular time, because their local
truncation errors are below or above both thresholds.  Alternatively,
the two simulations may have entirely different subdomain
boundaries in a given region at some other time.  Furthermore, AMR's decisions
exhibit hysteresis. Despite these issues,
most of our simulations do
exhibit convergence
with AMR tolerance, as shown in Sec.~\ref{sec:waveQuality}.

\subsection{Black hole masses, spins, and centers}
\label{sec:eval-mass-spins}

\subsubsection{Quasi-local definitions}
\label{sec:definitions}

Defining mass and spin in general relativity is nontrivial; for
recent reviews, see \cite{Szabados:2009eka, blanchet2011mass,
  Proceedings:2015hya}. The simplest definitions apply to asymptotically flat
spacetime, in the limit of approaching either spacelike infinity $i^{0}$
(i.e.~the ADM mass and angular momentum~\cite{Arnowitt:1962hi})
or future null infinity $\mathscr{I}^{+}$ (the Bondi mass and angular
momentum~\cite{Bondi:1962px}). These ADM and Bondi quantities
give global values for the entire spacetime, including binding energy and energy in gravitational waves, but they do not yield the masses and
spins of the individual component black holes.
One possible approach for determining component masses and spins,
while the black holes are far from merger, is
to perform asymptotic matching in a buffer region~\cite{Pound:2015tma}.
This approach, however,
has not been explored in numerical simulations, and becomes invalid as
the black holes approach merger.

Instead, NR simulations rely on ``quasi-local'' mass and
spin~\cite{Szabados:2009eka} measurements from AHs.  Quasi-local masses and spins recover the Kerr mass and spin when evaluated on the AH of a Kerr black hole, and also evolve in agreement with tidal torquing
and heating approximations~\cite{Scheel:2014ina}.
The definition of quasi-local mass that we adopt relies on our chosen
measure of quasilocal spin.
Given an apparent horizon $\mathcal{H}$ within the current
constant-time hypersurface $\Sigma$, and a
vector field $\phi^{i}$ tangent to $\mathcal{H}$,
the component of spin angular momentum inside this 2-surface
generated by $\phi^{i}$
is given by \cite{Ashtekar:2003hk} (following the conventions
of~\cite{Lovelace:2008tw})
 \begin{align}
  \label{eq:S_phi_def}
  S_{\phi} \equiv \frac{1}{8\pi} \int_{\mathcal{H}}
  \phi^{i} s^{j} K_{ij} \, dA \,.
\end{align}
Here $s^{i}$ denotes the outgoing spacelike unit normal to $\mathcal{H}$
tangent to $\Sigma$, $K_{ij}$ is the extrinsic
curvature of $\Sigma$ (with sign conventions given in Appendix~\ref{app:signs}), and $dA$ is the induced proper area element
on $\mathcal{H}$.

In axisymmetry, making $\phi^{i}$ the symmetry's
rotational Killing vector field yields the corresponding
conserved angular momentum. But black holes in a binary merger are only approximately axially
symmetric long before merger and after ringdown. With no exact rotational
Killing vector available, instead we
follow Refs.~\cite{Cook:2007wr,Lovelace:2008tw}
and compute approximate Killing vectors on the apparent horizon.
For details on our procedure
for measuring quasi-local spin, and
a discussion of other approaches, see~\cite{Owen:2017yaj}; here, we
briefly summarize our method.

We solve an eigenvector problem to find the
three tangent vectors $\{\phi_{(1)}^{A}, \phi_{(2)}^{A}, \phi_{(3)}^{A}\}$
that come closest to solving the Killing equation.
Here, capital Latin indices $A, B, \dots$ run over the two-dimensional tangent
space of $\mathcal H$.
The Killing equation implies that a Killing vector is divergence-free, which means
the vector has vanishing expansion and shear.
We choose to begin with expansion-free vectors
$\phi_{(i)}^{A} = \epsilon^{AB} D_{B} z_{(i)}$, writing them
as the ``curl'' of a potential $z_{(i)}$. Here $\epsilon^{AB}$ and $D_{A}$\
are the induced Levi-Civita tensor and covariant derivative of $\mathcal{H}$,
respectively. Minimizing the average squared shear of $\phi^{A}$ over the
surface then yields an eigenvector problem for the three $z_{(i)}$.
As in Ref.~\cite{Lovelace:2008tw}, we fix
the normalization of each $z_{(i)}$ by requiring the
variance of each $z_{(i)}$ to agree with the same variance in
the Kerr metric (cf. Eq.~(A22) and the surrounding discussion
in~\cite{Lovelace:2008tw} for details).
Finally, we define the spin magnitude $S$ as the Euclidean magnitude of
a vector with three components of angular momentum found by inserting
 the three $\phi_{(i)}^{A}$ with smallest eigenvalues into
 Eq.~(\ref{eq:S_phi_def}):
\begin{align}
  \label{eq:S_total_def}
  S \equiv \sqrt{S_{\phi_{(1)}}^{2}+S_{\phi_{(2)}}^{2}+S_{\phi_{(3)}}^{2}}
  \,.
\end{align}

We define the ``spin function'' $\Omega \equiv
\epsilon^{AB}D_{A}\omega_{B}$, where $\omega_{A}$ is the projection of
the 1-form $K_{ij}s^{j}$ into the tangent space of $\mathcal{H}$.
Then, we take the three moments of the spin function~\cite{Owen:2017yaj} to compute
the direction of the spin,
\begin{align}
  \label{eq:chi_hat_Om}
  \hat{\chi}_{\Omega m}
  \equiv \frac{1}{N}
  \int_{\mathcal{H}} \vec r \, \Omega  \, dA \,,
\end{align}
where $\vec r$ is the Euclidean position vector in the coordinate
system of the simulation.
We choose the normalization factor $N$ so that the Euclidean
norm of $\hat{\chi}_{\Omega m}$ is 1.  Finally, we define the full
dimensionful spin vector as
\begin{align}
  \label{eq:S_def}
  \vec S \equiv S \hat{\chi}_{\Omega m} \,.
\end{align}

With the dimensionful spin
in hand, we define the mass interior to $\mathcal{H}$ from the
Christodoulou formula for (uncharged) Kerr BHs~\cite{Christodoulou:1972kt},
\begin{align}
  \label{eq:M_chr}
  M^{2} \equiv M_{\textrm{irr}}^{2} + \frac{S^{2}}{4 M_{\textrm{irr}}^{2}} \,,
\end{align}
where the irreducible mass $M_{\rm irr}$
depends only on the apparent horizon's area:
\begin{align}
  \label{eq:M_irr}
  M_{\mathrm{irr}}^{2} \equiv \frac{A}{16\pi} = \frac{1}{16\pi} \int_{\mathcal{H}} dA \,.
\end{align}
Even though the Christodoulou relation is, strictly speaking, only justified
for stationary BHs, we also use it on dynamical AHs as a quasi-local mass.

From the dimensionful spin vector $\vec S$ given by Eq.~\eqref{eq:S_def} and the mass
$M$ given by Eq.~\eqref{eq:M_chr}, we define the dimensionless spin vector
\begin{align}
  \label{eq:chi_def}
  \vec \chi \equiv \frac{\vec S}{M^{2}} \,.
\end{align}
We compute the magnitude of this vector $\vec \chi$ using the
Euclidean norm. Equation~\eqref{eq:M_chr} implies $0 \leq |\vec \chi| < 1$ for all of our black holes. For a
discussion of how well this relation is satisfied in simulations of merging
black holes with nearly extremal spins, see Ref.~\cite{Lovelace:2014twa}.

Finally, we define a coordinate center $\vec x$ for each AH as the
surface-area weighted average of the location of the AH,
\begin{align}
  \label{eq:AHCenter}
 \vec x = \frac{1}{A} \int_{\mathcal H} \vec r \, dA \,.
\end{align}
This means that the (irreducible) mass dipole moment of the surface vanishes
in a coordinate system centered on $\vec x$.
In practice, we require that this condition be only approximately satisfied by truncating
our representation of the shape of the AH at $\ell = 1$ in an expansion in spherical harmonics,
as shown in Eqs.~(37)--(40) in~\cite{Hemberger:2012jz}.

\subsubsection{Definition of relaxation time and reference time
for measuring initial BH quantities}
\label{sec:defin-relax-time}

The initial data do not perfectly describe
two black holes in quasi-equilibrium.  At the start of each
simulation, therefore, the geometry relaxes to
equilibrium on the dynamical time-scale of the individual BHs, changing
the masses and spins of each BH by a fractional amount of order
$10^{-5}$, and emitting a spurious pulse of gravitational radiation
(often referred to as ``junk radiation'').  Our simulations do not
attempt to resolve this pulse of short-wavelength gravitational
radiation. Therefore, BH quantities (like their masses) fluctuate with an
amplitude of about $\sim 10^{-5}$ for a few $100M_0$, before the
fluctuations subside.  Here, $M_0$ denotes the sum of the two
Christodoulou masses at $t=0$. Subsequently, BH quantities vary on the
inspiral time-scale.

To avoid the impact of junk radiation on our output quantities,
we define a reference time
$t_{\rm ref}>0$ as early in the
simulation as possible,\footnote{In principle, one can
choose any time to define the BBH parameters, since the dynamics
provide a unique map from one time to the next.  In practice, for
analytical understanding and comparison with post-Newtonian theory, we
want to choose the earliest possible time, where the post-Newtonian
approximation is most accurate.} but after the initial transients have decayed.
We define a relaxation time $t_{\rm relax}>0$ at which we deem
that transients have decayed, and then we set $t_{\rm ref}$ to be
at least $t_{\rm relax}$.
We extract the ``initial'' BH properties we ascribe to each black hole
in our simulations at $t_{\rm ref}$.
For most simulations in the catalog $t_{\rm ref}=t_{\rm relax}$,
but we allow the two times to be different because they represent
different concepts and because for some simulations (e.g. comparisons
with waveform models or other NR codes) we desire to
specify parameters at particular times.
Since we do not attempt to resolve the initial transients, waveforms
computed at different resolutions correspond to
binaries with slightly different physical parameters.  Because of our
high precision, these small differences complicate our
convergence testing, as discussed further in
Sec.~\ref{sec:waveQuality}.
We also recommend to only use the
gravitational waveforms for retarded time
$u>t_{\rm ref}$.

In practice, we compute $t_{\rm relax}$ as follows:
We begin by defining a window of size $T_\mathrm{window} = 300M_0$ and
a time interval $\delta t = 10M_0$.  Considering the time series
$M_{\rm irr,1}(t)$ of irreducible masses for the primary black hole,\footnote{%
In some earlier simulations, we used the timeseries of the magnitude
of the dimensionless spin.  For most simulations, these two
quantities give comparable relaxation times.  In practice, the areal
mass tends to oscillate less than the spin magnitude.%
}
we compute its
standard deviation $\sigma_{n}$ for sliding time windows $t\in
[n\delta t, n\delta t+T_{\rm window}]$, $n=0, \ldots 30$.  We compute
a running average of the $\sigma_{n}$ using sequential sets of $10$
segments.  As the junk radiation propagates away, the running average
of $\sigma_{n}$ decreases with time.  We identify the earliest time at
which the running average stops decreasing, calling the center of this
time window $t_\text{relax,1}$ (if this condition fails, we set
$t_{\rm relax, 1}=600M_0$).  We repeat this calculation for the secondary BH,
and take the larger of the two values as the final relaxation time,
$t_{\rm relax}=\max(t_{\rm relax, 1}, t_{\rm relax, 2})$.

\subsubsection{Remnant masses, spins, and recoil velocities}
\label{sec:remnant-masses-spins}

After merger and ringdown, there remains a single remnant BH with its own mass,
spin, and a recoil velocity (a ``kick'' caused by asymmetry in momentum
carried by GWs).

As described in Sec.~\ref{sec:definitions},
our simulations extract at regular intervals
the BH mass $M(t)$, spin $\vec\chi(t)$ and center $\vec x(t)$ from local quantities on each apparent horizon.
This is also true after merger, when we extract the mass, spin, and center 
for the AH of the remnant from local quantities
on the common apparent horizon.
These quantities are highly dynamical immediately after merger because the remnant
horizon is strongly, dynamically curved when it forms. After the common AH forms, it relaxes, quickly
at first, and then rings down over a characteristic
timescale determined by the remnant's spin~\cite{Berti:2005ys,Berti:2009kk}.
After merger, we compute the remnant AH mass and
spin on a dense (though not uniformly sampled) set of
times. To compute the final AH mass and
spin, we simply split these time samples into thirds, and choose the mass and spin averaged over the final third of the time samples.
This approach seeks to mitigate small, residual
time-dependent variations
in the remnant mass and spin caused by numerical noise.

We also employ a simple procedure to estimate the \emph{coordinate}
recoil velocity (this turns out to be very close to the well-defined
recoil velocity arising from gravitational-wave momentum
flux integrals~\cite{Ruiz:2007yx, Gerosa:2018qay}).
We compute the coordinate center of the remnant AH as we do for the individual AHs during inspiral, using Eq.~(\ref{eq:AHCenter}).
Taking the last third of time
samples of $\vec x(t)$, we model each of its components 
with a least-squares fit to a linear function of time.
We then interpret the slopes of these fits as the coordinate velocities
of the remnant BH.

\subsection{Gravitational wave extraction}
\label{sec:grav-wave-extr}

We extract the emitted gravitational waves from our simulations through two independent
methods. The first computes the Newman-Penrose scalar $\Psi_4$ on a
set of coordinate spheres centered at the initial (coordinate) center
of mass of the two black holes. This is done by computing the Weyl
tensor, projecting with a flat-space
orthonormal null tetrad to form $\Psi_4$, and expanding in terms of spherical
harmonics of spin weight $-2$.  See Refs.~\cite{Pfeiffer:2007yz, Boyle:2007ft} for
details. We do not use a properly orthonormalized null
tetrad in computing $\Psi_4$ from the Weyl tensor, nor do we 
use anything other than coordinate-sphere extraction
surfaces. Therefore, our computation of $\Psi_4$ at a finite radius
differs from the standard definition by a multiplicative factor
$1+\mathcal{O}(1/r)$. We eliminate these differences by extrapolating
the waveforms to future null infinity (discussed in Sec.~\ref{sec:extrapolation})
to remove these and other near-zone
effects. We also remove some artifacts of our choice of coordinates in the initial
data via center-of-mass correction (discussed in Sec.~\ref{sec:com-correction}).

Applications using the waveforms from the SXS catalog should use the extrapolated waveforms with center-of-mass corrections; however, for diagnostic purposes, we also make available
the raw, finite-radius
spherical-harmonic modes $r\Psi_4^{\ell,m}$ (available in the SXS catalog
as files named
\texttt{rPsi4\_FiniteRadii\_CodeUnits.h5}).  In each case, the value of
$\Psi_{4}$ can be evaluated at a point $(\theta, \phi)$ using
\begin{equation}
  \label{eq:Psi4_at_a_point}
  \Psi_{4}
  =
  \sum_{\ell,m} \Psi_{4}^{\ell,m}\, {}_{-2}Y_{\ell,m}(\theta, \phi),
\end{equation}
where ${}_{-2}Y_{\ell,m}$ are the spin-weight $s=-2$ spherical
harmonics, using the conventions given in Ref.~\cite{Brown2007}.

The second gravitational-wave extraction method is independent of the
first one. In this method, we compute the metric perturbation directly using Sarbach
and Tiglio's~\cite{Sarbach:2001qq} formulation of the Regge-Wheeler
and Zerilli equations~\cite{Regge:1957td, Zerilli:1970se}.  First,
we compute the metric perturbation $\delta g_{ab}=g_{ab}-\eta_{ab}$
about a Minkowski background $\eta_{ab}$.
From the variables of our first-order formulation of Einstein's equations, we also read off the first time and spatial derivative of $\delta g_{ab}$.  We evaluate
$\delta g_{ab}$ and its derivatives on a
set of coordinate spheres centered at the initial (coordinate) center
of mass of the binary, and on each of these coordinate spheres we expand these
quantities 
in terms of spin-weighted spherical harmonics.
We then compute the spin-weighted spherical-harmonic modes of the Regge-Wheeler
quantity $\Phi^{(-)}$
and the Zerilli quantity $\Phi^{(+)}$, which are combinations of the
metric perturbations and their derivatives given in
Eqs.~(16--18), (22--29), and~(A12--A21) of
Ref.~\cite{Rinne:2008vn}.\footnote{We use the opposite sign convention for $\Phi^{(+)}$ as does Ref.~\cite{Rinne:2008vn};
in particular, we replace the minus sign in front of
Eq.~(29) of Ref.~\cite{Rinne:2008vn} with a plus sign.
Our sign convention agrees with that of Ref.~\cite{Ruiz:2007yx}, and ensures
that a linearized wave in TT gauge satisfies Eqs.~(\ref{eq:hplus_sign}) and~(\ref{eq:hcross_sign}) of Appendix~\ref{app:signs}, assuming Eqs.~(\ref{eq:WaveAmplitudesFromRWZScalars}) and~(\ref{eq:strain_at_a_point}).}
Finally, we compute the modes
of the strain using (cf.~Eq.~(83)
of~\cite{Nagar:2005ea} and Eq.~(4.34) of~\cite{Ruiz:2007yx})
\begin{equation}
  \label{eq:WaveAmplitudesFromRWZScalars}
  r\, h^{\ell,m} = \sqrt{(\ell-1)\ell(\ell+1)(\ell+2)}
    \; (\Phi^{(+)}_{\ell, m} + i \Phi^{(-)}_{\ell, m}).
\end{equation}
Note that $\Phi^{(\pm)}$ are not gauge invariant in the general sense,
but are sometimes referred to as such in the context of perturbation
theory~\cite{Moncrief1974, GerlachSengupta, Sarbach:2001qq,
  Martel:2005ir}.  Specifically, the definitions of $\Phi^{(\pm)}$
involve quantities that are invariant at first order under
infinitesimal gauge transformations about a fixed background, assuming
small metric perturbations from that background.  However, gauge
changes that cannot be treated as infinitesimal affect the waveforms
at a significant level.  In Sec.~\ref{sec:com-correction}, we describe
and remove some such gauge effects that are present in the waveforms
even after extrapolation to future null infinity.

For diagnostic purposes, we make the modes $r h^{\ell, m}$ of the raw finite-radius quantities available in the SXS
catalog as files \texttt{rh\_FiniteRadii\_CodeUnits.h5}. However, as with the
$\Psi_4$ waveforms described above, the raw finite-radius quantities contain
near-zone and gauge effects. For applications using the waveforms in the SXS catalog,
one should instead use the version of these waveforms that
are extrapolated to future null infinity and center-of-mass corrected
(see Sec.~\ref{sec:waveform-post-processing} below). The strain
at a point $(\theta, \phi)$ can be evaluated as
\begin{equation}
  \label{eq:strain_at_a_point}
  h
  =
  (h_{+} - i\, h_{\times})
  =
  \sum_{\ell,m} h^{\ell,m}\, {}_{-2}Y_{\ell,m}(\theta, \phi).
\end{equation}
Finally, note that each file in the SXS catalog describing raw, extrapolated, or
center-of-mass corrected $h$ contains the real and
imaginary parts of $h^{\ell, m}$,
which differs by a minus sign in the imaginary component from the format described in
Ref.~\cite{Brown2007}. More significantly, all waveforms in our catalog prior
to this release have had an overall sign change in the
definition of the strain. The strain as 
given in Eq.~\eqref{eq:strain_at_a_point} reflects the current definition of the
waveforms in our catalog, for more information see Appendix~\ref{app:signs}. For
details on checking the sign convention of the waveform files, see Appendix~\ref{sec:waveform-files}.

\subsection{Waveform post-processing}
\label{sec:waveform-post-processing}

Our catalog contains waveforms that are extrapolated to future null infinity 
and corrected for 
center-of-mass motion.
Here, we detail the extrapolation and
center-of-mass corrections applied to each $h$ and $\Psi_{4}$ waveform
in this catalog.

\subsubsection{Extrapolation}
\label{sec:extrapolation}

During the evolution, we extract each waveform at a
series of times on a set of concentric coordinate spheres surrounding the binary,
decomposed in modes of spin-weighted spherical harmonic
functions. We then extrapolate the waveforms to future null infinity,
$\mathscr{I}^{+}$.  Our method is similar to the one described
previously in Refs.~\cite{Boyle:2009vi,Taylor:2013zia}, but
we modify it to permit accurate extrapolation of precessing systems as follows.
We transform the waveform modes into a corotating
frame~\cite{Boyle:2013nka}, in which the rotation is factored out, so that the corotating waveform modes
vary slowly in time, even for precessing systems.
We then simply extrapolate the real and imaginary parts of the corotating waveform modes.
Ref.~\cite{Boyle:2009vi} shows that such slow time-varying behavior is crucial to
convergence of the extrapolation process.
We then transform the
extrapolated result back into an inertial frame.
Our previous extrapolation method~\cite{Boyle:2009vi,Taylor:2013zia} decomposed the complex waveform modes into phase and amplitude, which results in slow temporal variations \emph{only} for non-precessing binaries.

For each simulation, we compute both $h$ and $\Psi_{4}$ independently
(Sec.~\ref{sec:grav-wave-extr}).
We extrapolate both quantities to $\mathscr{I}^{+}$ by the same
method, though for simplicity we will only describe extrapolation of
$h$ here.
We extract the strain waveform as $h^{\ell, m}(T_{i}, R_{j})$
on a grid of coordinate times $T_{i}$ (where $T=0$ is the start
  of the simulation) and on coordinate spheres of fixed radii $R_{j}$.
The extraction radii are chosen between $R = R_{\rm min}$ and the outer
boundary of the simulation domain, typically with about 24 extraction
radii spaced uniformly in $1/R$.
For most simulations in the catalog, we choose
the innermost extraction radius $R_{\rm min}$ to be $100M_0$.
For newer simulations, we
chose $R_{\rm min}$ to be at least $\pi/\Omega_0$, where
$\Omega_0$ is the initial orbital frequency; thus $R_{\rm min}$ is at
least one gravitational-wave wavelength from the origin, and the waveform
there is not dominated by near-field effects.
We explicitly compute the areal radius of each sphere
$\rar_{j}$, which
depends on the time, by integration using the evolved
metric.  We also extract the average value of
the metric component $g^{TT}$ over this sphere.\footnote{%
  Each quantity
  used in the extrapolation is made available for download in files
  named \texttt{rh\_FiniteRadii\_CodeUnits.h5} and
  \texttt{rPsi4\_FiniteRadii\_CodeUnits.h5}, documented in
  App.~\ref{sec:format-hdf5-data}.}
These quantities allow
us to define the retarded time
\begin{equation}
  \label{eq:retarded_time}
  u_{i,j} = \mathlarger{\int}_{0}^{T_{i}} \sqrt{ \frac{-1/g_{j}^{TT}(T)}
    {1 - 2 \MADM / \rar_{j}(T)} } dT
  - \rar_{j}(T_{i}) - 2\MADM \ln \left[ \frac{\rar_{j}(T_{i})}
    {2\MADM} - 1 \right],
\end{equation}
where $\MADM$ is the ADM mass measured in the initial data.  The
second and third terms are essentially the familiar tortoise
coordinate of the Schwarzschild solution, while the first term is a
correction to the time coordinate.  We choose this retarded time so
that the one-form $du$ is approximately null with respect to the
evolved metric~\cite{Boyle:2009vi}.

Using these quantities, we can also express the waveform modes
$h^{\ell,m} (u_{i,j}, r_{i,j})$.  Then, we interpolate the
data to a common set of retarded times, $u_{k}$.  We construct
this
set to be the largest subset of $u_{0, j}$ such that the 
waveform at each
radius has
known values at each time $u_{k}$.  We also interpolate
the areal radius of each sphere to
the set of common times, so
that the waveform can now be expressed as $h^{\ell,m} (u_{k},
r_{k,j})$.

The next step is to rotate the waveform at each radius into a
corotating frame~\cite{Boyle:2013nka}.  To avoid the
complication of extrapolating the transformation to the corotating frame itself, we simply choose the outermost extraction radius to define the corotating
frame.  We cannot expect the waveform at any other radius to be in
\emph{precisely} its own corotating frame, but in practice we still achieve our
objective of ensuring that the waveform at each radius is slowly varying.
We denote the waveforms in this frame as
$\corotating{h}^{\ell,m} (u_{k}, r_{k,j})$.

Now, with slowly varying data tabulated on a common set of retarded
times and a series of radii, we can extrapolate the waveform to
infinite radius by approximating each mode with a polynomial of order
$N$:
\begin{equation}
  \label{eq:extrapolation}
  \corotating{h}^{\ell,m} (u_{k}, r)
  \approx
  \sum_{j=0}^{N}\frac{\corotating{h}^{\ell,m}_{(j)}(u_{k})} {\rar^{j+1}}.
\end{equation}
At each time step, we choose the coefficients $\corotating{h}^{\ell,m}_{(j)}$
to minimize the sum of the squared differences
between the numerical data at that time and the polynomial value---real and
imaginary parts being treated separately.  The asymptotic waveform in
the rotating frame is simply $\corotating{h}^{\ell,m}_{(0)} (u)$.  We then obtain
the final asymptotic waveform by inverting the rotation
that was applied above.

Note that the $m=0$ modes in SpEC
waveforms appear to be generally
unreliable, in the sense that they do not appear to converge with
increasing extrapolation order or varying extraction radii, and they
do not agree with CCE results~\cite{Taylor:2013zia}.  This is true
in the inertial frame 
for non-precessing systems, and typically true in
co-precessing~\cite{Hannam:2013oca, Pekowsky:2013ska, Boyle:2011gg} or
co-rotating~\cite{Boyle:2013nka} frames for precessing systems. This means that other modes in the inertial frame
may be polluted by these inaccurate co-rotating frame $m=0$ modes for precessing systems.

The code that we use to perform this entire extrapolation procedure is
available in the open-source python module
\texttt{GWFrames}~\cite{GWFrames}.
We provide all finite-radius
waveforms,
along with the extrapolated waveforms
for $N=2$, $3$, and $4$, for
both $h$ and $\Psi_{4}$.  The extrapolated data files also contain a
waveform from the outermost extraction radius, which is also
given as a function of the corrected retarded time $u$ and scaled by
the areal radius; we expect this will remove some (though not all) of
the gauge artifacts present in raw waveforms extracted at finite
radius. The waveform measured
at the outermost extraction radius is different from the waveform computed using $N=0$ in
Eq.~\eqref{eq:extrapolation}, because
the latter essentially averages
the contributions from data extracted at smaller radii---which can be
\emph{worse} than doing no extrapolation at all.
For this reason,
we do not provide a waveform extrapolated using
$N=0$, nor do we provide one for $N=1$~\cite{Boyle:2009vi}.

The choice of extrapolation order $N$ for a particular purpose must be
informed by the behavior of near-field effects.  If $\lambda$ is a
typical wavelength present in a given mode, we expect the higher-order
terms in the polynomial to scale not just as $1/\rar^{j}$ relative to
the lowest-order term, but as $(\lambda/\rar)^{j}$~\cite{Boyle:2009vi,
Thorne:1980ru}.  Thus, as the binary spirals in toward merger and the
length scales $\lambda$ become smaller, the polynomial will converge
much more quickly with $N$.  Thus, for example, if extrapolation with
$N=4$ is required for accurate results early in the waveform, then
$N=2$ may be sufficient closer to merger.  On the other hand, using a
large value of $N$ when the polynomial converges quickly can lead to
overfitting of features that are poorly modeled as functions of
retarded time and polynomials in $1/\rar$---so that $N=2$
extrapolation may actually be \emph{better} than higher-order
extrapolation during the merger and ringdown.  The general rule of
thumb, then, is to use higher-order extrapolation for analyses that
require more accuracy during the inspiral, and lower-order extrapolation for
analyses that
require more accuracy during the
merger and ringdown; there is
no single choice that is best for all applications.  In all cases, it
is best to test the dependence of results on the extrapolation order
by running an analysis multiple times using each of the various
extrapolation orders.
For the application of building waveform surrogate models~\cite[and
references therein]{Varma:2018mmi, Varma:2019csw}, we use $N=2$ for
the entire waveform.

\subsubsection{Center-of-mass correction}
\label{sec:com-correction}

The catalog provides extrapolated waveforms in two versions: one
without, and one with, a correction for displacement and drift of the
center of mass (COM).  These COM-corrected waveforms have filenames
ending in ``CoM''.  We recommend using COM-corrected waveforms for all
applications of our catalog, but provide the original data for
completeness and to allow for comparisons between COM-corrected and uncorrected waveforms.

These corrections are necessary because
the waveform modes depend on the origin of coordinates used to
define the spherical harmonics: if we move the origin,
the modes will be mixed.  Naively, we expect the origin to be centered
on the binary, as that is the natural choice and is used in
post-Newtonian models, for example.  That choice leads to desirable
features like relatively slowly varying mode amplitudes, and
frequencies that are roughly proportional to the orbital frequency
times the azimuthal number $m$ of the mode.\footnote{Spin-orbit
  coupling introduces effects for precessing systems where small
  additional components are present, and oscillate at frequencies
  roughly proportional to $(m \pm 1)$ times the orbital frequency.
  However, if the origin coincides roughly with the center of mass,
  these additional components can also be modeled to high accuracy
  without accounting for spurious mode
  mixing~\cite{Pekowsky:2013ska, Boyle:2014ioa}.} %
However, all binary-black-hole systems simulated with SpEC
contain essentially random offsets and drifts of the
origin of coordinates relative to the COM, causing mode-mixing
that manifests as irregular behavior in the waveform
modes.
To a good approximation, these
irregularities can be thought of as direction-dependent time
translations that appear uncorrelated
between different physical systems or different numerical
resolutions of the same
system.  These irregularities appear as essentially random
contributions to waveform modes
that are discontinuous with respect to changing physical
parameters.  These random and discontinuous effects would have to be
modeled by surrogate, EOB, and phenomenological waveform
models~\cite{Blackman:2017pcm, Varma:2018mmi, Blackman:2017dfb,
  Blackman:2015pia, Kumar:2018hml, Abbott:2016apu, Buonanno:1998gg, Bohe:2016gbl, Pan:2013rra,
  Khan:2015jqa, Hannam:2013oca, Damour:2007yf, Boyle:2008ge,
  Pan:2013tva, Kumar:2015tha, Kumar:2016dhh, Abbott:2016wiq}, or be optimized away in direct searches of detector
data~\cite{Heal:2017abq, Kumar:2013gwa, Lovelace:2016uwp,
  Lange:2017wki}.  By removing these effects, we simplify such
analyses.

Figure~\ref{fig:COMuncorrected} shows the translations and boosts of
the COM for two example simulations
(the non-precessing SXS:BBH:0314 and the precessing SXS:BBH:0627)
from our catalog, each shown at multiple resolutions.
We define the COM at each instant in time using the Newtonian definition,
\begin{equation}
  \label{COMdef}
  \vec{x}_{\txt{COM}} \equiv \frac{m_1(t)\vec{x}_1(t) + m_2(t)\vec{x}_2(t)}{m_{1}(t) + m_{2}(t)},
\end{equation}
where $m_1\mbox{ and } m_2$ are the Christodoulou masses (as defined in Sec.~\ref{sec:definitions})
of the primary and secondary black holes, and
$\vec{x}_1\mbox{ and } \vec{x}_2$ are the coordinate positions of the
AH centers as defined in Eq.~\eqref{eq:AHCenter}.
In our convention, black hole ``1'' is the more massive of the two.

Two causes contribute to the COM motion visible in Fig.~\ref{fig:COMuncorrected}.
First, while the initial data achieves $P_{\rm ADM}=0$~\cite{Ossokine:2015yla}, the initial transients
during relaxation to equilibrium may cause asymmetric GW emission, and thus impart a net linear momentum
onto the binary.  Second, since we do not attempt to resolve this junk radiation, we do not expect that the COM motion
in our simulations will be convergent; indeed, we observe essentially randomly varying coordinate velocities
of the COM for evolutions at different resolutions of the same initial data set.

\figCOMuncorrected

As can be seen in the right column plots of Fig.~\ref{fig:COMuncorrected}, the 
total COM displacement is generally only a fraction of the total mass $M_0$ of the
binary, and much smaller than the radii of the gravitational wave
extraction spheres ($R_j\gtrsim 100M_0$,
cf.~Sec.~\ref{sec:grav-wave-extr}).  Nevertheless, these small COM displacements do have a noticeable
impact on the higher-order modes of the computed gravitational radiation, as can be seen in the left column
plots in Fig.~\ref{fig:COMuncorrected}.
During inspiral and close to merger, the uncorrected higher-order waveform amplitudes oscillate---most
notably the $(l,m)$ modes $(2,1)$, $(3,1)$, and $(3,3)$.
This modulation is not expected on physical grounds; it is a gauge
effect caused by mode mixing that follows from the COM displacement.
The strongest effect of mode
mixing is the leaking of power from
the dominant $(2,\pm 2)$ modes into the subdominant modes, because we decompose our
waveform using spin-weighted spherical harmonics centered on an offset,
moving origin, which is unnatural.

We apply COM corrections to remedy these effects as a post-processing step. We
compute the parameters necessary for the correction from the simulation and, as the
corrections are BMS 
transformations~\cite{Bondi:1962px,Sachs:1962wk,Sachs:1962zza,Boyle:2015nqa},
they do not alter the physically meaningful aspects of the waveform.
Deciding how to correct waveforms for center-of-mass motion
is complicated and is described in a separate paper~\cite{Woodford:2019tlo}.
The final procedure itself is relatively straightforward, and
we summarize it here.

We implement the measurement of and corrections to the COM using the open-source python module
\texttt{scri}~\cite{scri, Boyle:2015nqa}.
While there is significant COM motion, as seen in Fig.~\ref{fig:COMuncorrected},
our COM correction deals only with the offset and drift---that is, the
linear motion.
Reference~\cite{Woodford:2019tlo} presents investigations into potential
physical contributions to the total COM motion, and the epicycles
seen in Fig.~\ref{fig:COMuncorrected}.

We remove these gauge effects using  translations and boosts.
To re-center the simulations, we first measure the offset and drift of the COM and then
retroactively apply the opposite motion to the 
waveform,
to cancel out that motion. We define a time
average of any quantity $Q(t)$,
\begin{align}
  \label{eq:time_ave}
  \ave{Q} \equiv \frac{1}{t_{f}-t_{i}}\int_{t_{i}}^{t_{f}} Q(t) \, dt\,.
\end{align}
We choose $t_i$
to be the relaxation time for the simulation (defined in Sec.~\ref{sec:defin-relax-time}) and set $t_f = 0.9t_{\rm CAH}$, where $t_{\rm CAH}$ indicates the time of common apparent horizon formation, so that
merger and ringdown are not included in the average.

Now we would like to find a translation $\vec{\alpha}$ and boost
$\vec{\beta}$ that give the best linear approximation to the motion of
the measured $\vec x_{\txt{COM}}$. Within the interval $[t_i, t_f]$ we perform a linear least-squares fit to $\vec x_{\rm COM}(t)$ resulting in a best-fit motion $\vec\alpha+\vec\beta t$.
As described in Appendix~E of \cite{Boyle:2015nqa},
the fit can be performed analytically, giving
\begin{subequations}
  \label{eq:COMsupertranslation_def}
  \begin{align}
    \vec{\alpha} = \frac{\ave{t^{2}}\ave{\vec{x}_{\txt{COM}}} - \ave{t}\ave{t\vec{x}_{\txt{COM}}}}{\ave{t^{2}}-\ave{t}^{2}}
    &= \frac{4(t_f^2 + t_ft_i+t_i^2) \ave{\vec{x}_{\txt{COM}}} - 6(t_f + t_i)\ave{t\vec{x}_{\txt{COM}}}}{(t_f - t_i)^2}
    \,, \\
    \vec{\beta} = \frac{\ave{t\vec{x}_{\txt{COM}}}-\ave{t}\ave{\vec{x}_{\txt{COM}}}}{\ave{t^{2}}-\ave{t}^{2}}
    &= \frac{12\ave{t\vec{x}_{\txt{COM}}} - 6(t_f + t_i)\ave{\vec{x}_{\txt{COM}}}}{(t_f - t_i)^2}
    \,,
  \end{align}
\end{subequations}
where the second equality of each line comes from
$\ave{t}=\frac{1}{2}(t_{i}+t_{f})$, $\ave{t^{2}} = \frac{1}{3}(t_{i}^{2}+t_{i}t_{f}+t_{f}^{2})$.

We then apply a displacement to negate the linear motion of the
COM given by $\vec{\alpha}+\vec{\beta}t$, computing this displacement
separately for each resolution
of each simulation. Reference~\cite{Boyle:2015nqa} first showed that this method
of COM corrections does indeed remove a large fraction of mode mixing and remedy
the COM offset and drift. Reference~\cite{Woodford:2019tlo} further confirms that
applying the COM correction does improve all waveforms in the SXS simulation catalog,
and introduces a robust and quantifiable method for this purpose.

\figCOMvals

Figure~\ref{fig:COMvals} shows translation, boost, and total displacement values
for spin-aligned (top row) and precessing (bottom row) simulations in the catalog. 
More recent simulations in the catalog use an 
improved initial-data method~\cite{Ossokine:2015yla} that achieves $P_{\rm ADM}=0$ 
in the initial data even for precessing systems and that reduces the overall 
displacement of the COM, especially for precessing cases.
For most systems $\vec{\alpha}$ and $\vec{\beta}t_{\rm CAH}$ are comparable.
Further details on the COM correction method and analysis can be found in~\cite{Woodford:2019tlo}.

\section{Parameter space coverage}\label{sec:paramSpace}

\figphysicalParamSpace

Expanding the catalog from the original 174 configurations to
  \Nconfigs{} configurations has
substantially improved our coverage of the BBH
parameter space. Figure~\ref{fig:physicalParamSpace} shows the
binary mass ratio $q=m_1/m_2 \ge 1$ and the dimensionless spin magnitudes $|\chi_1|$
and $|\chi_2|$ for the simulations in our catalog.
Each point in the scatter plots in Fig.~\ref{fig:physicalParamSpace}
represents a simulation, while the histograms show the relative number
of simulations with the given range of mass ratio and dimensionless
spin magnitudes.
The masses and spins plotted here are measured at the reference
time, as discussed in Sec.~\ref{sec:defin-relax-time}.
In the scatter plots, we see a substantial number of precessing simulations with mass ratios up to
$q=4$ and $|\chi_A|\leq 0.8$, which were produced in order to construct the
surrogate models of~\cite{Blackman:2017pcm, Varma:2018aht, Varma:2019csw}.
The subscript $A$ corresponds to the larger ($A=1$) and smaller
($A=2$) black holes.
In addition, we show
improved coverage of the nonprecessing subspace with mass ratios up to $q=8$
and $|\chi_A|\leq 0.8$.  New simulations in this part of the parameter space were produced in order to construct
the surrogate model of~\cite{Varma:2018mmi}.

In contrast, there remain large regions that are unexplored
in all BH merger catalogs, including ours.
The projections in $q-|\chi_{A}|$ space in Fig.~\ref{fig:physicalParamSpace} show
that while we have fairly dense coverage at low mass ratios, mass ratios larger than $q=4$ remain sparsely
explored or completely unexplored.
Similarly, aside from a few equal-mass, equal-aligned spin cases, the region of spin magnitudes
above 0.8 remains almost completely unexplored.
Few simulations exist with both high spins and high mass ratio.
These are especially challenging, as they require high resolution and delicate control
of the computational domain, including the shapes, sizes, and positions of the
excised regions inside the black-hole horizons (see, e.g.~\cite{Scheel:2014ina}).
For example, most simulations in the catalog with $q>4$ are non-precessing.

\figNorbits

Figure~\ref{fig:Norbits} shows a histogram summarizing the number of
orbits before merger
in our simulations.  Most simulations
have between 10 and 30 orbits.  This length is sufficient for many
gravitational-wave applications, particularly when the systems have
higher total masses, and thus remain in LIGO's  sensitive frequency
band for fewer orbits.
Spanning LIGO's sensitive frequency band for binaries with lower total mass
is more difficult. This can be done either by producing longer simulations, or through
hybridizing numerical simulations by attaching the final orbits to an approximate
post-Newtonian waveform to cover earlier times
(see, e.g.~\cite{MacDonald:2012mp} and the references therein).
In both cases, achieving sufficient accuracy for applications to gravitational wave
science remains challenging.

\figEccHist

Figure~\ref{fig:EccHist} shows a histogram of the estimated initial
orbital eccentricity for each of our simulations.  For most simulations,
we tune our initial data via an iterative
procedure~\cite{Pfeiffer:2007yz, Buonanno:2010yk, Mroue:2012kv} to produce
nearly quasicircular orbits; the orbital eccentricities of these
simulations are almost all below $6\times 10^{-4}$.
For some simulations we intentionally wish to study eccentricity so
we omit the eccentricity-reduction step; these can be seen as the
tail in Fig.~\ref{fig:EccHist}.
As described in Ref.~\cite{Mroue:2010re,Buonanno:2010yk}, we estimate orbital
eccentricity by a least squares fit of an analytic function to the
time derivative of the orbital frequency $d\Omega/dt$.  This fit
is performed during the $\sim 2$ orbits following $t_{\rm relax}$. It
captures the monotonic inspiral-driven long-term trend in
$d\Omega/dt$, and overlaid oscillatory variations caused by orbital
eccentricity [see Eqs.~(70) and~(76) of~\cite{Buonanno:2010yk}].  Motivated
by Eq.~(68) of~\cite{Buonanno:2010yk}, we report an eccentricity (effectively
averaged over the first two to three orbits by our fit) computed with
\begin{equation}\label{eq:e}
e_0 = \frac{B_\Omega}{2 \Omega_0 \omega_\Omega},
\end{equation}
where $B_\Omega$ and $\omega_\Omega$ represent the
amplitude and frequency of oscillations in $d\Omega/dt$, and $\Omega_0$
is the orbital frequency at time $t=0$.
We fit to $d\Omega/dt$ rather than to $\Omega$ because the time
derivative magnifies eccentricity-induced oscillations, making them
easier to fit when eccentricity is small.
At large
separation, Eq.~\eqref{eq:e} reproduces the Newtonian definition of
eccentricity to linear order in $e_0$.  Since we neglect higher-order
corrections in $e_0$, large values of $e_0$ reported for our simulations
are only rough estimates of the actual orbital eccentricity.
We have not made an effort to precisely recover any
post-Newtonian eccentricities~\cite{DamourDeruelle1985}, which we
expect to differ from Eq.~\eqref{eq:e} by fractional corrections of order
$(v/c)^2$.

\subsection{Coverage in spin space}
\label{sec:coverage-spin-space}

Of the seven dimensions of parameter space for quasi-circular mergers, six
are spin components.  This high dimensionality makes the parameter
space difficult to sample densely and uniformly.  Previous
catalogs~\cite{GaTechCatalog, RITCatalog} have discussed coverage in
spin space but without a thorough exploration of the degree of
coverage.

Visualizing the coverage is difficult due to the high
dimensionality, but we can focus on certain physically relevant
combinations of the spin parameters. A commonly used spin combination
that strongly affects total waveform phase is the
effective spin~\cite{Ajith:2009bn,Santamaria:2010yb,Hannam:2013oca},
\begin{align}
\chi_{\rm eff} \equiv \frac{(m_1\vec{\chi}_1 + m_2 \vec \chi_2)\cdot \hat{L}} { m_1+m_2} = \frac {m_1 \chi_{1\parallel}
+ m_2 \chi_{2\parallel}}{m_1+m_2}
\,.
\label{eq:chieff}
\end{align}
Here $\hat L$ is the direction of the instantaneous Newtonian orbital
angular momentum, and we carry out the projection of the spins onto
$\hat L$ using the Euclidean metric.  When the effective
spin is positive, the black holes merge more slowly, causing the
gravitational-wave frequency to increase more slowly; conversely,
when the effective spin is negative, the black holes merge more quickly,
causing the gravitational-wave frequency to increase more quickly.
Another benefit to considering
the effective spin is that while the spin directions can precess in
a complicated manner, $\chi_{\rm eff}$ is conserved up to at least the
2nd post-Newtonian order~\cite{Racine:2008qv}.
By contrast, the in-plane components
\begin{align}
  \vec{\chi}_{A\perp} \equiv \vec{\chi}_{A} - (\vec{\chi}_{A} \cdot \hat{L}) \hat{L}
\end{align}
are more relevant for recoil kicks~\cite{Gonzalez:2007hi,
  Campanelli:2007cga, Lousto:2011kp, Lousto:2012gt} and precession
dynamics~\cite{Kesden:2014sla, Gerosa:2015tea}.

\figchieffxy

Figure~\ref{fig:chieffxy} is one view of the distribution of black-hole
spins, measured at the reference time.
Shown are the effective spin $\chi_{\rm eff}$ and the magnitudes of the
in-plane vectors, $\chi_{A\perp} \equiv |\vec{\chi}_{A\perp}|$.
The catalog contains a large number of non-spinning and
aligned-spin simulations,
leading to a spike at low in-plane spins.
There is a population of simulations with $|\vec{\chi}_{A\perp}|
\approx 0.8$, which were used to build the surrogate models
of~\cite{Blackman:2017dfb, Blackman:2017pcm}.

\figprecAngles

A different view of the parameter space is relevant for understanding
precession dynamics.  Namely, a convenient combination of parameters
is given by the two ``tilt angles'' $\theta_{A\,L}$~\cite{Gerosa:2015tea},
\begin{align}
  \label{eq:costhetaAL}
  \cos\theta_{A\,L} \equiv \hat{\chi}_{A} \cdot \hat{L}
  \,,
\end{align}
and the in-plane angle $\Delta\Phi$ between the two
$\vec{\chi}_{A\perp}$ vectors,
\begin{align}
  \label{eq:DeltaPhi}
  \cos\Delta\Phi \equiv \hat{\chi}_{1\perp} \cdot \hat{\chi}_{2\perp}
  \,,
\end{align}
with the sign determined according to $\sgn\Delta\Phi
\equiv \sgn\{ \hat{L} \cdot [ (\chi_{1} \times \hat{L}) \cdot
(\chi_{2} \times \hat{L}) ]  \}$~\cite{Gerosa:2015tea}.
Figure~\ref{fig:precAngles} plots these three parameters.
Because of the large number of aligned-spin simulations, there is
pileup at values of $\cos\theta_{A\,L} = \pm 1$.  There are also
larger number of simulations with purely in-plane spins, leading to
another pileup at values $\cos\theta_{A\,L} = 0$.  The distribution in
$\Delta\Phi$ is relatively flat.

\subsection{Parameter space coverage and LIGO measurements}
\label{sec:LIGO_coverage}

\figLIGOcomparison

Figure~\ref{fig:LIGOcomparison} compares the parameter space coverage of the SXS catalog to selected
astrophysical measurements of coalescing black holes by the LIGO detectors, obtained from~\cite{Vitale:2017cfs,salvatore_vitale_2018_1313235}. The left panel illustrates the magnitudes of the
dimensionless spin vectors of the two black holes and their corresponding spin tilt angles $\theta_{AL}$
for each simulation in the catalog as measured at the reference time
(dots, color coded by the simulation's mass ratio).
These are plotted over the marginalized
posterior distributions for these quantities for GW151226~\cite{Abbott:2016nmj}, one of only two observed black hole binaries, to date,
with evidence for non-zero spin~\cite{LIGOScientific:2018mvr} (denoted by greyscale pixels).
The right panel shows the effective spin and the mass ratio of each
system in the catalog and,
for GW151226 and three additional gravitational wave
detections, the 90\% credible contours of the marginalized
2-dimensional posterior distributions.

In both cases, the SXS catalog covers a large part of the relevant parameter space that is
consistent with the LIGO measurements. In particular, LIGO observations point towards black-hole binaries with small effective spins, a region probed well with the majority of the SXS
simulations. However, the comparison also suggests that there are regions of the parameter
space where the coverage is sparse, especially for unequal mass systems. Future simulations
will help fill in this region of the parameter space and serve to improve the accuracy of
waveform models.

\section{Waveform quality}\label{sec:waveQuality}

As discussed in Sec~\ref{sec:SpECsum}, during a simulation SpEC
employs dynamical adaptive $p$- and $h$-refinement, adding
or subtracting grid points, or splitting or joining subdomains, according
to local measures of truncation error and a globally specified
AMR tolerance. The vast majority
of the simulations in the catalog presented here have been
run at multiple levels of this
AMR tolerance (which we will henceforth
call ``multiple resolutions'' for brevity).
As discussed in Sec~\ref{sec:SpECsum}, because of the adaptivity, a small
number of our simulations do not show convergence with resolution; these
are noted in Figs.~\ref{fig:levmismatch} and~\ref{fig:Psihmismatch}.

It is not always straightforward how to compare two waveforms. Some
comparison criteria,
such as phase or amplitude differences for a particular spherical-harmonic waveform mode, might be important for
certain applications and not for others.
Following \cite{Chu:2015kft,Blackman:2017dfb},
here we take an approach motivated by the practical
application of the waveforms to gravitational wave science.
We define an overlap $\mathcal O$ between waveforms and examine
overlaps between waveforms
from simulations with the same initial parameters but differing
resolutions.
This provides one measure for the accuracy of our waveforms.
A second accuracy measure involves comparing
waveforms extracted using the extrapolated metric perturbation $h$
to those extracted at the same resolution with the extrapolated
Weyl scalar $\Psi_4$.

To define the overlap $\mathcal O$, we employ a standard inner product
$\langle h^1 , h^2 \rangle$ between complex waveforms  $h^1$ and $h^2$ given by
\begin{align}
\label{eq:FDInnerProd}
\langle h^1 , h^2 \rangle = \mathlarger{\int}_{-\infty}^{+\infty} \frac{\tilde h^1 (f) \tilde h^2{}^*(f)}{S_n(|f|)} \, df \,,
\end{align}
where the tilde denotes a frequency domain signal.
This product is real if $h^1(t)$ and $h^2(t)$ are real.
The quantity $S_n(|f|)$ is the noise power spectral density,
which quantifies the spectrum of the colored Gaussian noise, and is
also used to whiten the signals in a gravitational wave detector.
Here, we assume a flat noise spectrum, and set $S_n(|f|)=1$.
This choice has the benefit that the results described here are
independent of the total mass of the binaries.
These overlaps are in qualitative agreement with those obtained using
the Advanced LIGO noise curve, when the total mass of the system is in
the range $50-100 M_{\odot}$, as shown in Fig.~4 of Ref.~\cite{Varma:2019csw}.
Details of our implementation of the inner product are
given in App.~\ref{app:mismatches}.

Given this inner product, we define an overlap that accounts for the
information in both
polarizations $h_+$ and $h_\times$ in $h(t,\theta,\phi)=h_{+}-i h_{\times}$, as if
measured by two ideally oriented detectors located at a given
$(\theta, \phi)$ position in the source frame,
\begin{align}
\label{eq:TwoDetOverlap}
\mathcal{O}(h^1,h^2) =
\text{Re}
\left[
\frac{\langle h^1, h^2 \rangle}{
    \sqrt{\langle h^1, h^1 \rangle
      \langle h^2, h^2 \rangle}}
\right]
\,.
\end{align}
For waveforms that are identical up to a re-scaling, $\mathcal{O}=1$,
otherwise $\mathcal{O}<1$.  We are interested in quantifying the
difference in numerical simulations of the identical physical system
at different numerical resolutions, so $h^1$ and $h^2$ correspond to
evolutions of the same initial data set, but with different numerical
resolution.  As described in detail in Appendix~\ref{app:mismatches},
when comparing two such waveforms $h(t, \theta,\phi)$ we allow
for an overall rotation $\delta\phi$ of one relative to the other, and a
time offset $\delta t$.  For each pair of waveforms and direction $(\theta,\phi)$ from the binary to the detector, this
results in a mismatch
\begin{equation}
\label{eq:Mismatch}
\mathcal M(h^1,h^2) = 1 - \max_{\delta \phi, \delta t}\mathcal{O}(h^1,h^2,\delta \phi, \delta t).
\end{equation}
For every configuration we include in
Figures \ref{fig:levmismatch} and \ref{fig:Psihmismatch},
we evaluate the mismatch at 20 distinct source frame directions
$(\theta,\phi)$.

\figNRMismatch

Figure~\ref{fig:levmismatch} shows a histogram of the resulting mismatches between the two
highest
resolution simulations for the \NsimsMoreThanOneRes{} simulations in the catalog with more than one resolution.
The majority of the simulations, \NsimsMoreThanTwoRes{}, have more than two resolutions and in these cases we can
assess
the convergence of the waveforms by comparing the mismatch between the two highest
resolutions
with that between the second and third highest resolutions.
We expect the former mismatch to be smaller than the latter mismatch in waveforms that
converge with
increasing resolution settings, and when this fails to occur we label the waveform
``nonconvergent.''

The top panel of Fig.~\ref{fig:levmismatch} depicts the mismatches for the extrapolated
metric
perturbation $h$ computed using the Regge-Wheeler-Zerilli extraction technique described
in Sec.~\ref{sec:grav-wave-extr}.
The bottom panel
depicts the mismatches between the Weyl scalar $\Psi_4$,
weighted by
$f^{-2}$ in order to give the same frequency weighting as $h$.
We see that overall the mismatches are small, with the mismatches broadly lying between
$\sim10^{-6}$ and $10^{-3}$, appropriate for many current applications to gravitational wave
science.
This is true for both the mismatches computed using $h$ and $\Psi_4 f^{-2}$.
There are \NmismatchesTotal{} mismatches plotted in each panel
of Fig.~\ref{fig:levmismatch}, corresponding to 20 detector directions $(\theta,\phi)$ for \NsimsTotal{} simulations; only
\NmismatchesNonconvergentH{} of these mismatches are nonconvergent as determined by $h$ and \NmismatchesNonconvergentPsi{} as determined by $\Psi_4f^{-2}$.

The error estimates in Fig.~\ref{fig:levmismatch} have a tail
  that extends to rather high mismatches. This feature was
  investigated in App.~B of Ref.~\cite{Varma:2019csw}, and shown to be
  a consequence of unresolved initial transients. As mentioned in
  Sec.~\ref{sec:defin-relax-time}, these initial transients cause
  simulations at different resolutions to correspond to binaries with
  slightly different physical parameters.  In particular, we find that
  the in-plane spins of different resolutions can be inconsistent with
  each other. Ref.~\cite{Varma:2019csw} used surrogate-modeling tools
  to show that the high-$\mathcal{M}$ tail in
  Fig.~\ref{fig:levmismatch} is dominated by the small differences in
  system parameters.  By training a surrogate model on high-resolution
  simulations, and evaluating it with spins of medium-resolution
  simulations, Ref.~\cite{Varma:2019csw} found mismatches (between the
  surrogate and medium-resolution NR waveforms) to always be below
  $\lesssim10^{-2}$.  Thus our error estimate is overly conservative
  and does not reflect the actual truncation error of the simulations.
  We expect that the actual truncation error tail in
  Fig.~\ref{fig:levmismatch} should only extend to $\lesssim 10^{-2}$
  rather than $\sim 10^{-1}$.

Furthermore, the mismatches in Fig.~\ref{fig:levmismatch} are
somewhat pessimistic measures of the accuracy of the waveforms, since
they actually measure the error in the second highest resolution, not
the highest.
Given the refinement scheme in SpEC, we cannot use extrapolation of the
convergent
waveforms to provide an error measurement on the highest resolution simulation.
The mismatches also combine all possible sources of error
together,
so that we cannot distinguish truncation error from other sources of error, such as
inaccuracies in
our prescription for extrapolating the waveforms to infinity.

\figPsihmismatch

Figure~\ref{fig:Psihmismatch} provides another perspective on the accuracy of the
waveforms in the catalog.
The top panel provides a histogram of the mismatches between $\ddot h f^{-2}$ and $\Psi_4
f^{-2}$
from for the highest resolution for each waveform.
This mismatch is between two quantities that are equal in theory, and so it provides
an independent assessment of the numerical accuracy of the waveform, and especially of the
two
waveform extraction techniques.
The factor $ f^{-2}$ ensures the same frequency weighting for the signals used in
the mismatches of both Figs.~\ref{fig:levmismatch} and~\ref{fig:Psihmismatch}.
The overall level of the mismatches is
lower here than in Fig.~\ref{fig:levmismatch}, with fewer cases extending beyond a mismatch
of $10^{-3}$.
Instead of comparing $\ddot{h}$ vs.~$\Psi_{4}$,
one could perform mismatches of $h$ against $\iint
\Psi_{4}$, but by using the time derivative $\ddot{h}$ we avoid the
difficulty of needing to fix two integration constants, which
is not straightforward~\cite{Reisswig:2010di}.

The bottom panel of Fig.~\ref{fig:Psihmismatch} is a scatter plot
of the mismatch between $\ddot h f^{-2}$ and $\Psi_4 f^{-2}$ at
the highest resolution and the same mismatch at the second-highest
resolution.  The dashed line has unit slope and helps to quickly
assess which mismatch is larger.  While we see a few more points above
the line, the plot indicates that the mismatch
between $\ddot h f^{-2}$ and $\Psi_4 f^{-2}$ is roughly independent of
resolution, but the scatter is wide with many outliers.
Furthermore, the histogram in the top panel of
Fig.~\ref{fig:Psihmismatch} is visually unchanged when computed with
the second-highest resolution rather than the highest resolution (as
is shown).  Therefore Fig.~\ref{fig:Psihmismatch} shows an additional
source of error, perhaps caused by differences in waveform extraction and 
extrapolation, which is independent of numerical resolution
and is smaller on average than numerical truncation error.
This also suggests numerical truncation error
affects the strong-field evolution more than it does the
wave propagation in the far zone; otherwise the different wave
extraction methods would show larger differences with numerical
resolution.

\figExtrapOrderMismatch

All waveforms in Figs.~\ref{fig:NR_Mismatch} and~\ref{fig:Psihmismatch} have been extrapolated to infinity as described in Secs~\ref{sec:extrapolation}.  To quantify errors in our extrapolation procedure, Fig.~\ref{fig:ExtrapOrderMismatch} shows mismatches between waveforms that are identical except for details of extrapolation. In particular, we compare the highest-resolution waveform from each of our simulations, extrapolated to infinity using extrapolation order $N=2$ (the standard choice we make, e.g., in surrogate models~\cite{Varma:2019csw}),
versus the same waveform extrapolated to infinity using extrapolation order $N=3$.  The mismatches peak below $\mathcal{M}\sim 10^{-6}$ with a tail that
extends to $10^{-3}$, demonstrating that by
this measure, errors in our extrapolation procedure are on average
unimportant compared to numerical truncation error.

A sufficient condition for two waveform models to be indistinguishable
is~\cite{Flanagan:1997kp, Lindblom:2008cm, McWilliams:2010eq, Chatziioannou:2017tdw}
\begin{align}
\label{eq:mismatchSNR}
\mathcal{M} < \frac{D}{2\rho^2},
\end{align}
where $\mathcal{M}$ is the mismatch [cf.~Eq.~\eqref{eq:Mismatch}]
 and $\rho$ is the signal-to-noise ratio (SNR) of the observation the models are
describing. Here $D$ is the number of relevant model parameters, with $D=8$ for spin-precessing systems.
For $\rho=24$ (the case for GW150914~\cite{Abbott:2016blz}), this corresponds
to a mismatch of $7 \times 10^{-3}$.
Comparing to Fig.~\ref{fig:NR_Mismatch}, we find that for most of
our simulations, different numerical resolutions are indistinguishable
in this sense, using a flat noise curve. This gives us confidence that our numerical waveforms are well suited for interpreting gravitational-wave observations as loud as
GW150914, provided that the total mass is sufficiently high that the waveform
is long enough to span the observed signal. However, note that using Eq.~(\ref{eq:mismatchSNR}) to determine whether two waveforms are indistinguishable in a particular detector would require using that detector's noise curve in the mismatch calculations and choosing a total mass for each mismatch.

\section{Remnant properties}\label{sec:remnant}

The remnant properties of the final black hole are of great interest
both astrophysically and for constructing semi-analytical waveform
models.  On the astrophysical side, the final mass, spin and kick
velocity give important information about the possible progenitors of
the system and may help distinguish BBH formation channels~\cite{Gerosa:2018wbw}. In
waveform modelling, the final mass and spins are important ingredients
in constructing the full waveform, determining the quasi-normal
modes whose superposition creates the ringdown signal. This requires
one to connect the parameters of the black holes far from merger to
those of the remnant.  Thus the task of inferring remnant properties
has
received a lot of attention~\cite{Herrmann:2007ex, Campanelli:2007cga,
    Gonzalez:2007hi, Gonzalez:2006md, Campanelli:2007ew,
    Rezzolla:2007rz, Rezzolla:2007rd, Kesden:2008ga,
    Tichy:2008du, Lousto:2007db, Barausse:2009uz,
    Pan:2011gk, Barausse:2012qz, Lousto:2012su,
    Lousto:2012gt, Healy:2014yta, Zlochower:2015wga,
    Hofmann:2016yih, Gerosa:2015tea, Jimenez-Forteza:2016oae,
    Healy:2016lce, Healy:2018swt, Varma:2018aht}.

In this section, we compare fits for the final mass and spin
magnitude from the literature to the simulations in the catalog, restricting our
attention to cases where the measured eccentricity is $\le 2\times 10^{-3}$. We
make use of the publicly available implementation of the fits in
LAL~\cite{LALsuite,LALsuite:nrutils}.
To estimate the error in NR data, we use the difference between the highest and
second highest resolution, where more than one resolution is available. We
define the errors in the fitted mass as $\Delta m = m_{\txt{NR}}-m_{\txt{fit}}$ and
similarly for the magnitudes of the final spin.

We begin by considering the fits for the remnant mass from Healy and
Lousto~\cite{Healy:2016lce} (HL2016) and Jiménez-Forteza et
al.~\cite{Jimenez-Forteza:2016oae} (UIB2016). As can be seen in
Fig.~\ref{fig:Mass_fits}, we find good agreement between fits and the NR
simulations. The errors in the fits are below
$|\Delta m| \lesssim 0.004 M$
for 90\% of the cases, more than an order of magnitude larger than the NR errors.
There is also a tail that extends
to larger negative errors which shows that the fits systematically overestimate the final
mass.

\figMassfits

We next consider the final dimensionless spin magnitudes, using the models from
Hofmann et al.~\cite{Hofmann:2016yih} (HBR2016) in addition to HL2016 and
UIB2016, as shown in Fig.~\ref{fig:Spin_fits}. For precessing cases we follow
current LIGO/Virgo analyses and evolve the spins (using the 3.5
post-Newtonian spin evolution equations) from the relaxed time to the
Schwarzschild innermost stable circular orbit (ISCO). Projections of these
spins along the Newtonian orbital angular momentum direction at ISCO are
used as inputs for the remnant mass and spin fits. The HL2016 and UIB2016 spin fits are augmented by the sum of the in-plane spin components at ISCO (for details see \cite{
  mcdaniel2016}). For all models, we find errors of order
$\Delta|\chi| \lesssim 0.01$ for 90\% of the cases,
more than an order of magnitude larger than the NR errors. While there is little
difference in the magnitude of the errors between different models, we find that
the HBR2016 model shows the least skew around 0, most likely because of special
correction factors included in the model to handle precessing cases.

\figSpinfits

More accurate fits~\cite{Varma:2018aht, Varma:2019csw} for the remnant mass, spin and recoil
kick velocity have recently been developed by training directly against some of
these simulations. The errors in these fits are comparable to the NR errors,
but the fits have been trained only
against simulations with $q\leq4$,
$\chi_1,\chi_2 \leq 0.8$.
However, they are shown to extrapolate reasonably to higher mass ratios and spins in~\cite{Varma:2018aht, Varma:2019csw}.
Note that when applying aligned-spin fits to precessing systems,
there is an ambiguity as to what time or frequency the precessing spins are to be 
evaluated when using the aligned-spin model. 
The fits in~\cite{Varma:2018aht, Varma:2019csw} resolve this ambiguity by training directly against precessing simulations.
Ref.~\cite{Varma:2018aht} also suggests that fits built only from
aligned-spin NR simulations can become inadequate at SNRs $\sim 5$ times that
of GW150914, so there is a need to calibrate directly to precessing
simulations.

\section{Conclusion}\label{sec:conclusion}

In this paper, we have presented a substantial expansion of the SXS catalog
of numerical-relativity simulations of black-hole binaries, which is publicly available
at \url{https://www.black-holes.org/waveforms}~\cite{SXSCatalog}. Our catalog now includes \Nconfigs{}
simulations (including \Nprec{} that are precessing) with a median length of
\medNGWcyc{} cycles of $\ell=m=2$ gravitational waves.
We have considerably expanded
our coverage of the parameter space, especially for mass ratios
up to 4 and spin magnitudes up to 0.8. While our catalog does include
simulations with mass ratios up to \qmax{} and simulations spin magnitudes up
to \chimax{}, the parameter space of high mass ratios, high spins,
or both remains highly challenging and largely unexplored both by our catalog and
by other catalogs. The remnant masses and spins agree well with existing
fits in the literature, although differences between the fits and our simulations
are larger than differences between our different resolutions; as a
result, improved fits have recently been constructed directly from NR 
simulations~\cite{Varma:2018aht}.

We also have assessed the quality of our numerical waveforms. We find that
mismatches between waveforms at different numerical resolutions are smaller
than $10^{-3}$ for the vast majority of simulations in our catalog, although
a few simulations have larger mismatches between different
resolutions.
As discussed in Sec.~\ref{sec:waveQuality}, current simulations
are adequate for parameter estimation with Advanced LIGO and
Virgo signals detected at the current level of sensitivity.
Significantly
louder observations, possible with
future ground- and space-based detectors,
might require numerical
waveforms with significantly higher accuracy.

In the future, we will continue to expand our catalog, with the ultimate goal of fully
covering the parameter space of binary black holes. This will likely require novel approaches
to enable simulations with both high mass ratios and nearly extremal, precessing spins. We
will also work toward longer simulations, which span detectors' frequency bands down to
smaller total masses, and more thoroughly explore cases with higher eccentricity. We will develop
improved initial data with less spurious ``junk'' gravitational
radiation (e.g.\ \cite{Varma:2018sqd}).
Finally, we are working towards computing gravitational waves using
Cauchy-characteristic extraction~\cite{Bishop:1996gt, Babiuc:2010ze,
  Taylor:2013zia, Handmer:2015dsa, Handmer:2016mls} rather than extrapolation.

\ack
We are pleased to thank
Davide Gerosa and
Josh Smith
for helpful discussions.
This work was supported in part by
the Sherman Fairchild Foundation, NSF
grants PHY-1708212 and PHY-1708213 at Caltech,
NSF grants PHY-1606654 and DGE-1650441 at Cornell,
NSF grants PHY-1606522, PHY-1654359, and AST-1559694 and by Dan Black and Goodhue-McWilliams
at Cal State Fullerton.
S.\ E.\ Field is partially supported by NSF grant PHY-1806665.
H.\ Fong and C.\ J.\ Woodford acknowledge support from NSERC of Canada
grant PGSD3-504366-2017, and H.~Fong further acknowledges support from
the University of Toronto and the Japan Society for the Promotion of
Science.
P.\ Schmidt acknowledges support from the NWO Veni grant no.\ 680-47-460.
This work used the Extreme
Science and Engineering Discovery Environment (XSEDE), which is supported by
National Science Foundation grant number ACI-1548562.  This research is part of
the Blue Waters sustained-petascale computing project, which is supported by
the National Science Foundation (awards OCI-0725070 and ACI-1238993) and the
state of Illinois. Blue Waters is a joint effort of the University of Illinois
at Urbana-Champaign and its National Center for Supercomputing Applications.
Computations were performed on
the GPC supercomputer at the SciNet HPC Consortium~\cite{Scinet}; SciNet is funded by: the Canada Foundation for Innovation (CFI) under the auspices of Compute Canada; the Government of Ontario; Ontario Research Fund (ORF) – Research Excellence; and the University of Toronto.
Computations were performed on the supercomputer Briare{\'e} from the Universit{\'e} de Montr{\'e}al, managed by Calcul Qu{\'e}bec and Compute Canada. The operation of these supercomputers is funded by the Canada Foundation for Innovation (CFI), NanoQu{\'e}bec, RMGA and the Fonds de recherche du Qu{\'e}bec-Nature et Technologie (FRQ-NT).
Computations were performed on the
Wheeler cluster at Caltech, which is supported by the Sherman Fairchild
Foundation and by Caltech; on NSF/NCSA Blue Waters under allocation NSF
PRAC--1713694; and on XSEDE resources Bridges at
the Pittsburgh Supercomputing Center, Comet at the San Diego Supercomputer
Center, and Stampede and Stampede2 at the Texas Advanced Computing Center,
through allocation TG-PHY990007N.
Computations were performed on the Minerva high-performance computer
cluster at the Max Planck Institute for Gravitational Physics in
Potsdam.

\appendix

\section{Contents of the SXS catalog}
\label{sec:sxs-catalog-contents}

The SXS catalog contains of a set of binary black hole simulations,
each labeled by an identification string of the form ``SXS:BBH:dddd'',
where ``dddd'' is a four-digit number.  A given catalog entry labeled
by ``SXS:BBH:dddd'' usually contains data from several resolutions,
i.e., several simulations with identical code and identical parameters
except for the AMR tolerance (cf.~Sec.~\ref{sec:SpECsum}).  These
resolutions are labeled ``LevN'', where ``N'' is an integer that
increases with finer resolution.  The resolution labels for different
catalog entries are not necessarily comparable. Each resolution starts
from identical initial data, sampled onto the appropriate grid for
that resolution.

The entries ``SXS:BBH:dddd'' are labeled roughly (but not exactly)
chronologically, so that larger numbers usually correspond to later
simulations with more recent versions of SpEC.  Sometimes two
different catalog entries have identical physical parameters (black
hole masses and spins), but are separate entries because they follow a
different number of orbits, have different orbital eccentricities or
outer-boundary radius, or because they use different prescriptions for
initial data or gauge conditions.
In addition, a number of simulations have been repeated using identical
or nearly identical parameters with an improved version of SpEC, 
including the aligned-spin simulations of \cite{Chu:2015kft}.
Once an entry is in the catalog, it
is never replaced by a newer simulation with the same ``SXS:BBH:dddd''
label; instead, the newer simulation is assigned a new label.  An
existing entry is modified only rarely, in the case of simple
corrections that can be done by postprocessing: for example, we found
that some of our waveforms in the catalog had the opposite overall
sign convention as intended, so the offending waveforms were changed
appropriately in the catalog and the version numbers of the modified
files were updated.

\subsection{Available data for each simulation}
\label{sec:available-data-each}

Each simulation in the catalog contains a \texttt{metadata.json} file
that lists physical and code parameters, derived quantities such as
remnant properties, and informational fields.
This file is documented in Table~\ref{metadata-fields} of
Sec.~\ref{sec:metadata-format}.

Each simulation also includes files that contain the
spherical-harmonic modes of gravitational waveforms, as described
in Sec.~\ref{sec:format-hdf5-data} and listed in Table~\ref{tab:h5-files}.
Finally, each simulation includes files \texttt{Horizons.h5} containing
masses, spins, and other properties of the apparent horizons.

All dimensionful quantities are given in arbitrary units.  All vector
and tensor quantities are given in an asymptotically Cartesian
coordinate system in which the black holes at the initial time are
along the $x$-axis (BH 1 at positive $x$), and the initial Newtonian
orbital angular momentum is along the $z$-axis.

\subsection{Metadata format}
\label{sec:metadata-format}

\newcommand{\firstgroupname}[1]{\multicolumn{3}{l}{\hspace{2em} #1}\\
\hline\noalign{\smallskip}}
\newcommand{\groupname}[1]{\noalign{\smallskip}\hline\firstgroupname{#1}}
\newcommand{\arrOrStr}{\parbox[t]{3.2cm}{Array of String,\\ or String}}

{
  \newlength{\colAdj}
  \setlength{\colAdj}{0pt}
  \addtolength{\tabcolsep}{\colAdj}
\begin{center}
  \begin{longtable}{@{\hspace*{0em}}p{5.7cm}lp{7.8cm}@{\hspace*{0em}}}
    \caption[Names, types, and descriptions of meanings of fields in
    \texttt{metadata.json} files.]{Names, types, and descriptions of
    meanings of fields in \texttt{metadata.json} files.}
    \label{metadata-fields} \\
    \hline
    Field name(s) & Type & Description \\
    \hline \hline\noalign{\smallskip}
\endfirsthead
\multicolumn{3}{c}%
{{\bfseries \tablename\ \thetable{} -- continued from previous page}} \\
    \hline
    Field name(s) & Type & Description \\
    \hline \hline\noalign{\smallskip}
\endhead
\noalign{\smallskip}
\hline \hline \multicolumn{3}{r}{Continued on next page} \\ \hline
\endfoot
\noalign{\smallskip}
\hline \hline
\endlastfoot

\firstgroupname{Identification}
		{\footnotesize \texttt{simulation\_name}} & string & A SXS-assigned identifier of the simulation \\
		{\footnotesize \texttt{alternative\_names}} & string[ ] & Comma-separated list of alternative names, longer, more descriptive, and/or indicating the specific series of simulations this configuration belongs to.  One of these alternative names is the `SXS:BBH:dddd' id-number, which is guaranteed to be unique.
\\
		{\footnotesize \texttt{keywords}} & string[ ] & Deprecated
\\
		{\footnotesize \texttt{point\_of\_contact\_email}} & string & Contact information for questions
\\
		{\footnotesize \texttt{authors\_emails}} & string[ ] & Deprecated
  \\
\groupname{References}
		{\footnotesize \texttt{simulation\_bibtex\_keys}} & string[ ] &
References which should be cited if this simulation is used
\\
		{\footnotesize \texttt{code\_bibtex\_keys}} & string[ ] & Deprecated \\
{\footnotesize \texttt{initial\_data\_bibtex\_keys}} & string[ ] & Deprecated
\\
{\footnotesize\texttt{quasicircular\_bibtex\_keys}} & string[ ] & Deprecated
\\
\groupname{Input parameters for initial data}
		{\footnotesize \texttt{object\{1,2\}}} & string & Keyword description to identify the object type.  One of \{\texttt{bh}, \texttt{ns}\}
  \\
		{\footnotesize \texttt{initial\_data\_type}} & string &
  Type of initial data.  One of\\
  && \texttt{BBH\_CFMS} -- conformally flat, maximal slice\\
  && \texttt{BBH\_SKS} -- superposed Kerr-Schild\\
		{\footnotesize \texttt{initial\_separation}} & double &
Coordinate separation $D_0$ between centers of compact objects, as passed to the initial data solver~\cite{Cook:2004kt,Buonanno:2010yk,Ossokine:2015yla}
\\
  {\footnotesize\texttt{initial\_orbital\_frequency}} & double &
Initial orbital frequency $\Omega_0$ passed to the initial-data solver~\cite{Buonanno:2010yk,Ossokine:2015yla}
\\
	{\footnotesize \texttt{initial\_adot}} & double &
Radial velocity parameter $\dot{a}_0$ passed to the initial data solver~\cite{Buonanno:2010yk,Pfeiffer:2007yz}
\\
\groupname{Measurements of initial data}
	{\footnotesize \texttt{initial\_ADM\_energy}} & double &
ADM energy of the initial data
\\
{\footnotesize \texttt{initial\_ADM\_linear\_momentum}} & double[3] &
ADM linear momentum of the initial data
\\
{\footnotesize \texttt{initial\_ADM\_angular\_momentum}} & double[3] &
ADM angular momentum of the initial data
\\
	{\footnotesize \texttt{initial\_mass\{1,2\}}} & double &
  Christodoulou mass of the apparent horizon of each body in initial data (Eq.~(\ref{eq:M_chr}); code units)\\
{\footnotesize\texttt{initial\_dimensionless\_spin\{1,2\}}}
& double[3] &
  Dimensionless spins of the BHs in the initial data (cf. Sec.~\ref{sec:eval-mass-spins})
  \\
{\footnotesize \texttt{initial\_position\{1,2\}}} & double[3] & Initial coordinates of the center of each body
\\
\groupname{Reference quantities}
{\footnotesize \texttt{relaxation\_time}} & double &
Time at which we deem junk radiation to have sufficiently decayed (code units)
\\
{\footnotesize \texttt{reference\_time}} & double &
Time at which the quantities are extracted from the evolution (code units)
\\
{\footnotesize \texttt{reference\_mass\{1,2\}}} & double &
  Christodoulou masses of the black holes at reference time, cf. Eq.~(\ref{eq:M_chr}) (code units)
\\
{\footnotesize \texttt{reference\_dimensionless\_spin\{1,2\}}}
& double[3] &
BH spins at reference time (cf.~Sec.~\ref{sec:definitions})
\\
	{\footnotesize \texttt{reference\_position\{1,2\}}} & double[3] &
Coordinates of the centers of the two bodies at reference time
\\
  {\footnotesize \texttt{reference\_orbital\_frequency}} & double[3] &
Orbital angular frequency vector at reference time
\\
{\footnotesize \texttt{reference\_mean\_anomaly}} & double &
Mean anomaly at reference time
\\
 {\footnotesize \texttt{reference\_eccentricity}} & double &
 Orbital eccentricity at reference time~\cite{Mroue:2010re}
  \\
\groupname{Merger/remnant quantities}
	{\footnotesize \texttt{number\_of\_orbits}} & double &
Number of orbits until formation of a common apparent horizon
\\
{\footnotesize \texttt{common\_horizon\_time}} & double &
Evolution time at which common horizon is first detected
\\
{\footnotesize \texttt{remnant\_mass}} & double &
Final mass of the remnant black hole after merger, cf.~Sec.~\ref{sec:remnant-masses-spins}
\\
  {\footnotesize\texttt{remnant\_dimensionless\_spin}} & double[3] &
Spin of the remnant black hole after merger, cf.~Sec.~\ref{sec:remnant-masses-spins}
\\
	{\footnotesize \texttt{remnant\_velocity}} & double[3] &
Linear velocity of the remnant black hole after merger, cf.~Sec.~\ref{sec:remnant-masses-spins}
\\
\pagebreak
\groupname{Code information}
	{\footnotesize \texttt{metadata\_version}} & int &
  Version of the metadata format itself. All simulations in this
  release of the catalog have version number 1\\
	{\footnotesize \texttt{spec\_revisions}} & string[ ] &
  (Array of) git revisions of the evolution code\\
	{\footnotesize \texttt{spells\_revision}} & string[ ] &
  (Array of) git revisions of initial data solver\\
\end{longtable}
\end{center}
\addtolength{\tabcolsep}{-1\colAdj}
}

\subsection{Format of the HDF5 data files}
\label{sec:format-hdf5-data}

\subsubsection{Waveform files}
\label{sec:waveform-files}
We provide several files in HDF5 format that contain extracted
gravitational waves.  Users should generally use
\textbf{\texttt{rhOverM\_Asymptotic\_GeometricUnits\_CoM.h5}} or
\textbf{\texttt{rMPsi4\_Asymptotic\_GeometricUnits\_CoM.h5}}, which
contain our best-effort gravitational wave modes, extrapolated to
future null infinity (cf. Sec.~\ref{sec:extrapolation}) and corrected
for center-of-mass effects (cf. Sec.~\ref{sec:com-correction}).  These
files contain several groups (i.e., the HDF5-equivalent of folders),
each one of them containing a separate set of GW waveform modes,
corresponding to different extrapolation order.  The groups are named
\texttt{Extrapolated\_N<int>.dir}, where the integer \texttt{<int>}
indicates the polynomial order of extrapolation.
See the discussion in Sec.~\ref{sec:extrapolation} regarding how to
choose extrapolation order.
In addition, there is a group \texttt{OutermostExtraction.dir} which
contains the GW modes at the largest available extraction radius,
without extrapolation, but scaled and corrected as described in
Sec.~\ref{sec:extrapolation} for consistency with the extrapolated
waveforms.

Each of these HDF5 groups contains one HDF5 dataset for each
$(\ell,m)$ mode; this dataset is named
\texttt{Y\_l<int1>\_m<int2>.dat}.  The first integer \texttt{<int1>}
indicates the value of $\ell$ for this particular mode, and the second
integer \texttt{<int2>} indicates the value for $m$.  \texttt{<int1>}
is always $\ge 2$, whereas $\texttt{<int2>}$ takes negative values for
negative $m$.  Each \texttt{Y\_l<int1>\_m<int2>.dat} dataset contains
three columns---either
\begin{align}
  \frac{u}{M}
  && \frac{r}{M}\text{Re} \left\{ h^{\ell,m} \right\}
  && \frac{r}{M}\text{Im} \left\{ h^{\ell,m} \right\}
    \intertext{or}
    \frac{u}{M}
  && r M \ \text{Re} \left\{ \Psi_{4}^{\ell,m} \right\}
  && r M \ \text{Im} \left\{ \Psi_{4}^{\ell,m} \right\}.
\end{align}
Here $u/M$ is the retarded time defined in
Eq.~\eqref{eq:retarded_time}, made dimensionless by division with the
sum of the two Christodoulou
  masses, $M=m_1+m_2$, where each mass is measured at the
reference time via Eq.~\eqref{eq:M_chr}.
The time spacing is non-uniform, with more points in regions of
higher GW frequency.  Note that the $h$
files contain the real
and imaginary parts of $h^{\ell, m}$ as opposed to the polarizations
$h_{+}$ and $h_{\times}$; see Eq.~\eqref{eq:strain_at_a_point}.

In addition to the primary files
\texttt{rhOverM\_Asymptotic\_GeometricUnits\_CoM.h5} and
\texttt{rMPsi4\_Asymptotic\_GeometricUnits\_CoM.h5}, some files with
intermediate GW data are provided.  These files contain the GW modes
without COM-correction and/or without extrapolation as listed in
Table~\ref{tab:h5-files}.  These files may sometimes be useful for
debugging purposes, but they should not generally be used.  Some of
these extra waveform-files contain extrapolated GW modes
(\texttt{extrap=yes}) in the same structure and format as just
described for \texttt{rhOverM\_Asymptotic\_GeometricUnits\_CoM.h5} and
\texttt{rMPsi4\_Asymptotic\_GeometricUnits\_CoM.h5}.  The files with
non-extrapolated waveforms (\texttt{extrap=no} in
Table~\ref{tab:h5-files}) contain groups for different extraction
radii, named \texttt{Rxxxx.dir}, where the four-digit integer
\texttt{xxxx} indicates the radius of the extraction sphere.  SpEC
chooses extraction radii always at integer values, so there is no
rounding in this number.  However,  note that the radius is given in
dimensionful code units without division by $M$.

Some of the HDF5 
files will have an HDF5 dataset called
\texttt{VersionHist.ver} in the root HDF5 group, which stores the entire
version history of the file. If a
file does not have this dataset,
then it is on version 0. \texttt{VersionHist.ver} is an array
of pairs where the first element in the pair is the git commit id for
the parent of the commit responsible for the change, and the second
element is a description of the change
from the previous
version. 

Only version 1 of \texttt{rh*.h5} files follows the sign convention for the
strain $h$ in Eqs.~(\ref{eq:hplus_sign},\ref{eq:hcross_sign}). For version 0,
there is an overall minus sign in Eqs.~(\ref{eq:hplus_sign},\ref{eq:hcross_sign}),
and hence an overall sign difference between the
waveforms contained in version 0 and version 1 of the \texttt{rh*.h5} files. Notice that this also implies that relation between $\ddot{h}$ and
$\Psi_4$ in Eq.~\eqref{eq:psi4_h} is off by a sign for \texttt{rh*.h5} files on version 0.
We make our best effort to ensure each type of data file is on the same version across
all runs, but we still recommended checking the version when working with files across
different runs.

\begin{table}
\begin{tabular}{l|cccc}
Filename & time & data & extrap? & COM? \\
\noalign{\smallskip} \hline \noalign{\smallskip} \hline \noalign{\smallskip}
\texttt{\textbf{rhOverM\_Asymptotic\_GeometricUnits\_CoM.h5}} & $u/M$ & $\frac{r}{M}h^{\ell,m}$   & \yes & \yes \\
\texttt{\textbf{rMPsi4\_Asymptotic\_GeometricUnits\_CoM.h5}} & $u/M$ & $rM\psi_4^{\ell,m}$   & \yes & \yes \\
\noalign{\smallskip} \hline \noalign{\smallskip}
\texttt{rhOverM\_Asymptotic\_GeometricUnits.h5}      & $u/M$ & $\frac{r}{M}h^{\ell,m}$   & \yes & \no  \\
\texttt{rMPsi4\_Asymptotic\_GeometricUnits.h5}       & $u/M$ & $rM\psi_4^{\ell,m}$       & \yes & \no  \\
\noalign{\smallskip} \hline \noalign{\smallskip}
\texttt{rh\_FiniteRadii\_CodeUnits.h5}               & $T$   & $r\,h^{\ell,m}$           & \no  & \no  \\
\texttt{rPsi4\_FiniteRadii\_CodeUnits.h5}            & $T$   & $r\,\psi_4^{\ell,m}$      & \no  & \no  \\
\noalign{\smallskip} \hline \noalign{\smallskip} \hline \noalign{\smallskip}
\end{tabular}
\caption{\label{tab:h5-files} %
  Data files that contain GW modes. The first two entries are the
  preferred ones; all other files should be used only when a clearly
  understood need arises.  The ``time'' and ``data'' labels indicate the contents
  of the time and data columns, where $u$ denotes the retarded time,
  corrected for finite-radius effects via
  Eq.~\eqref{eq:retarded_time}, and $T$ is the raw unnormalized time
  coordinate of the underlying NR simulation.  The last two columns
  indicate whether extrapolation and center-of-mass corrections were
	applied [cf. Secs.~\ref{sec:extrapolation} and~\ref{sec:com-correction}].}
\end{table}

\subsubsection{Apparent horizon files}

For each simulation, we provide one file, \texttt{Horizons.h5},
containing data computed from
the apparent horizons.
Each file contains 3 groups
named \texttt{AhA.dir}, \texttt{AhB.dir}, and \texttt{AhC.dir}; these
groups correspond to the two individual horizons (labeled ``A'' and ``B'')
and the common apparent horizon (labeled ``C''). Typically, horizon A
corresponds to the black hole with the larger initial Christodoulou mass.

Table~\ref{tab:horizons-groups} lists the seven HDF5 datasets given
for each horizon (i.e., for each HDF5 group).
The first column of each data set contain the
time $T$ from the simulation, and the next column (for scalar quantities)
or next three columns (for spatial vector quantities) contain the data. Quantities with
dimension are each given in the same arbitrary units, and all vector quantities
are given in the (asymptotically) inertial frame of the simulation.
Note that the times $T$ are not spaced uniformly.

\begin{table}
\begin{tabular}{l|cccl}
Dataset name& columns & time & data\\
\noalign{\smallskip} \hline \noalign{\smallskip} \hline \noalign{\smallskip}
\texttt{ArealMass.dat} & 2 & $T$ & $M_{\rm irr}$ & Eq.~(\ref{eq:M_irr})\\
\texttt{ChristodoulouMass.dat} & 2 & $T$ & $M$ & Eq.~(\ref{eq:M_chr})\\
  \texttt{CoordCenterInertial.dat} & 4 & $T$ & $\vec{x}$ & Eq.~\eqref{eq:AHCenter} \\
\texttt{DimensionfulInertialSpin.dat} & 4 & $T$ & $\vec{S}$ & Eq.~(\ref{eq:S_def})\\
\texttt{DimensionfulInertialSpinMag.dat} & 2 & $T$ & $S$ & Eq.~(\ref{eq:S_phi_def})\\
\texttt{chiInertial.dat} & 4 & $T$ & $\vec{\chi}$ & Eq.~(\ref{eq:chi_def})\\
\texttt{chiMagInertial.dat} & 2 & $T$ & $|\vec{\chi}|$ & (Euclidean norm)\\
\noalign{\smallskip} \hline \noalign{\smallskip} \hline \noalign{\smallskip}
\end{tabular}
\caption{\label{tab:horizons-groups}
The contents of each group of the
\texttt{Horizons.h5} file; these datasets are provided for each of the
individual apparent horizons and the common apparent horizon.}
\end{table}

\section{Computation of mismatches}
\label{app:mismatches}

One way to assess the quality of numerical waveforms is to compute
mismatches between waveforms that are supposed to be equal but are
computed using different numerical resolution parameters, boundary
conditions, extraction procedures, or methods of extrapolating to
infinity.  Our method of computing mismatches is similar but not
identical to the procedure described in Appendix D
of~\cite{Blackman:2017dfb}.

We begin with two waveforms in the frame of our simulation
\begin{align}
h^1(t,\theta,\phi) &= h_+^1(t,\theta,\phi) - i h_\times^1(t,\theta,\phi),\\
h^2(t,\theta,\phi) &= h_+^2(t,\theta,\phi) - i h_\times^2(t,\theta,\phi),
\end{align}
where the complex waveforms include both gravitational-wave
polarizations, and the angular dependence of the waveforms is usually
written as a sum of spin-weighted spherical-harmonic modes.  We
define $t=0$ as the time of maximum power in the waveform, and we
truncate the first $500 M$ of the waveform to eliminate effects
of initial transients sometimes known as ``junk radiation.''
This $500 M$ is uniform across all waveforms, distinct from the
reference time discussed in Sec.~\ref{sec:defin-relax-time}, which
can vary across resolutions.

If two waveforms $h^1$ and $h^2$ differ only by an overall coordinate
rotation or an overall time shift, we would like the two waveforms to
compare as equal, and therefore to have an overlap of unity.  We
accomplish this through two steps.  First, before we compute the
mismatch, we rotate both waveforms so that the orbital angular
momentum lies along the $+z$ axis at some fiducial time $t_0$.  In
other words, at $t=t_0$ the coordinate frame is momentarily aligned
with the minimally rotating coprecessing frame of
Ref.\cite{Boyle:2011gg}.  We choose $t_0 = t_{\rm begin}+1000M$, where
$t_{\rm begin}$ is the earliest time that is covered by both waveforms,
after the above-described truncation of the first $500M$ of each waveform
(if we are comparing three waveforms, as is the case
in Fig.~\ref{fig:levmismatch} where we compute overlaps between three
numerical resolutions, we choose $t_{\rm begin}$
to be the earliest time that is covered by all three waveforms).
Second, after the rotation, we allow one of the
waveforms (choose it to be $h^2$) to have an arbitrary azimuthal angle
shift and an arbitrary time shift: $h^2(t,\theta,\phi) \to h^2(t+\delta
t,\theta,\phi+\delta \phi)$.  The shift $\delta \phi$ corresponds to a
redefinition of the position of the two black holes in the orbit at
$t=t_0$.  Later we will minimize the mismatch over $\delta t$ and
$\delta\phi$.

When evaluating the accuracy of our numerical waveforms, we do not
wish to ignore the polarization information contained in these
waveforms. Furthermore, while we are interested in the angular
dependence of the waveforms, we do not want to concern ourselves with
antenna patterns of detectors. Therefore, we compute overlaps assuming
the most optimistic detector scenario: an ideal network of two
detectors located at $(\theta, \phi)$ relative to our source frame
and oriented normal to the direction of wave propagation, one
detector measuring $h_+(t)$ and the other measuring $h_\times(t)$.
This motivates the two-detector overlap defined in
Eq.~\eqref{eq:TwoDetOverlap} (see also Appendix D of~\cite{Blackman:2017dfb}).

To compute the Fourier transforms in Eq.~\eqref{eq:FDInnerProd} we
use an FFT after tapering the ends of the time-domain waveforms.
For the window function we use a Planck-taper window (Eq.~(7) of~\cite{McKechan:2010kp}).
This function depends on four parameters $t_1,t_2,t_3,t_4$: it
rises smoothly from zero at  $t=t_1$ to unity at $t=t_2$, and
falls smoothly from unity at $t=t_3$ to zero at $t=t_4$.  We
choose $t_1=t_{\rm begin}$ and $t_4=t_{\rm end}$, where $t_{\rm end}$ is
the latest time that is covered by both waveforms (or all three, in the
case of Fig.~\ref{fig:levmismatch}).  We choose
$t_2$ to be the time of the 10th zero-crossing
of the real part of the $(2,2)$ mode after $t=t_1$, and we choose
$t_3$ to be $50M$ after the peak of the waveform.
Before we compute the transforms, we pad with zeros and interpolate
each resolution's time-domain waveform onto the same evenly spaced set
of time samples, where the number of samples is chosen to be
\begin{align}
  \log_{2 }N_{\text{resamp}} = 1+\lceil\log_{2} \max_{\text{LevN}} N_{\text{samp}} \rceil
  \,.
\end{align}
We then truncate the Fourier transforms at a low-frequency cutoff $f_{\rm min}$
chosen to be
twice the waveform angular velocity (as defined by Ref.~\cite{Boyle:2013nka})
at $t=t_2$,
and a high-frequency cutoff $f_{\rm max}$ chosen to be
16 times the waveform angular velocity at the time of peak waveform power; the extra factor of 8 is chosen to resolve
up to $m=8$ spherical-harmonic modes, with an extra margin of a factor of 2.

The optimization over $\delta t$ can be simplified by noting that
the Fourier transform of $h(t+\delta t)$ is
$\tilde{h}(f)e^{2\pi i\delta t}$, so the overlap takes the form
\begin{align}
\mathcal{O}(\delta \phi, \delta t)
    ={}& \mathrm{Re}\left[\frac{\langle h^1, h^2(\delta t, \delta \phi)\rangle}{
        \sqrt{\langle h^1, h^1\rangle \langle h^2(\delta t, \delta \phi), h^2(\delta t,\delta
\phi)\rangle}}\right] \notag \\
    ={}& \mathrm{Re}\left[\frac{1}{
        \sqrt{\langle h^1, h^1\rangle \langle h^2(\delta \phi), h^2(\delta \phi)\rangle}}
            \int\frac{\tilde{h}^{1}(f)\tilde{h}^{2*}(f; \delta \phi)}{S_n(|f|)}
            e^{-2i\pi\delta t} df \right].
    \label{eq:final_overlap}
\end{align}
To compute $\max_{\delta t}\mathcal{O}(\delta \phi, \delta t)$ for a fixed $\delta \phi$, we
evaluate the integral in Eq.~(\ref{eq:final_overlap}) efficiently for many values of $\delta
t$ simultaneously using an inverse FFT, and we take the maximum value.  We then use standard
numerical maximization techniques to maximize over $\delta \phi$, resulting in the mismatch
$\mathcal M$ defined in Equation~\eqref{eq:Mismatch}.

In order to include the effect of higher-order spherical-harmonic
modes, we evaluate the mismatch at 20
points on the unit sphere, evenly spaced in $\cos\theta$ and $\phi$,
that describe the direction of the detector with
respect to the source. The
mismatch computed using each of those
20 directions
is plotted separately in Figs.~\ref{fig:levmismatch} and~\ref{fig:Psihmismatch}.

Note that some mismatch computations also explicitly minimize over a
polarization-angle shift $\psi$, which rotates the polarization tensor
that we use to decompose the waveform into the two polarizations $h_+$
and $h_\times$.  On the $z$ axis, optimization over $\psi$ is
  precisely degenerate with optimization over $\delta\phi$, even when
  all modes are included~\cite{Boyle:2016tjj}; off the $z$ axis this
  degeneracy is broken.  Here we consider $h^1$ and $h^2$ to be
different even if they differ only by a polarization-angle
shift, since we are considering the case of a detector network that
measures both polarizations, and since our numerical waveforms contain
polarization information. Hence we do not minimize over a
polarization-angle shift when computing overlaps and mismatches.

\section{Sign conventions}
\label{app:signs}

With so many sign conventions in the literature, we explicitly provide
an outline of sign conventions used in SpEC. Here, Greek indices represent
four-dimensional spacetime coordinate indices, and Latin indices represent
three-dimensional coordinate indices for a space-like hypersurface. For
a spacetime metric $\psi_{\mu\nu}$ with signature $(-,+,+,+)$, we foliate the
spacetime into space-like slices orthogonal to a timelike unit one-form $t_\mu$,
\begin{equation}
  t_\mu = - N\, \nabla_\mu t \,,
\end{equation}
where $t$ is a scalar function representing global time, and $N$ is the lapse.
With a shift vector $N^i$, we define the spatial metric and the extrinsic
curvature, respectively,
\begin{align}
  g_{\mu\nu} &= \psi_{\mu\nu} + t_\mu t_\nu \,, \\
  K_{\mu \nu} &= - \frac{1}{2} \mathcal{L}_{\mathbf{t}} g_{\mu \nu} \,, \\
  K_{ij} &= \frac{1}{2N} \left ( -\partial_0 g_{ij} + N^k \partial_k g_{ij}
            + 2g_{k(i} \partial_{j)} N^k \right ) \,,
\end{align}
where $K_{ij}$ represents the spatial components of the extrinsic curvature, and
the subscript $0$ indicates the time component.
While the sign convention for $K_{ij}$ is rather varied in the literature, the
one chosen here follows the Misner-Thorne-Wheeler convention and is found in
many prominent texts~\cite{GravitationMTW, Frolov:1998wf,
  Baumgarte:2010ndz, Shibata:2016nr, Rezzolla:2013rh}.
There are also several texts that follow the opposite sign
convention~\cite{Wald:1984gr, Carroll:2004sg, Poisson:2004tool}.

We define the 4-volume form $\epsilon_{\alpha\beta\gamma\delta}$ and the 3-volume
form
$\epsilon_{ijk}$ on the spatial slices as follows,
\begin{align}
  \epsilon_{0123} &= |\text{det}(\psi_{\mu\nu})|^{1/2} \,, \\
  \epsilon^{0123} &= -|\text{det}(\psi_{\mu\nu})|^{-1/2} \,, \\
  \epsilon_{ijk}  &= t^\mu \epsilon_{\mu ijk} \label{eq:3volumeform} \,, \\
  \epsilon_{123}  &= |\text{det}(g_{ij})|^{1/2}
                    \,,
\end{align}
and all others related by complete antisymmetry, $\epsilon_{abcd} =
\epsilon_{[abcd]}$.
Note that some texts define the 3-volume form as
$\varepsilon_{ijk} = \epsilon_{ijk\mu} t^\mu$, which incurs a minus sign
relative to the definition in Eq. (\ref{eq:3volumeform}), so that
$\varepsilon_{ijk} = -\epsilon_{ijk}$. 

We define the Christoffel symbols and the Riemann, Ricci, and Weyl tensors,
respectively, following the Misner-Thorne-Wheeler convention,
\begin{align}
  \Gamma_{\alpha \beta}^{\gamma} &= \frac{1}{2}\psi^{\gamma \lambda}\left(\partial_\beta
\psi_{\lambda\alpha}
                    + \partial_\alpha \psi_{\lambda\beta}
                    - \partial_\lambda \psi_{\alpha \beta} \right) \,, \\
  R^{\alpha}{}_{\beta\gamma\delta} &= \partial_\gamma \Gamma^{\alpha}_{\delta\beta} -
\partial_\delta \Gamma^{\alpha}_{\gamma\beta}
                + \Gamma^{\alpha}_{\gamma \lambda} \Gamma^{\lambda}_{\delta\beta}
                - \Gamma^{\alpha}_{\delta \lambda} \Gamma^{\lambda}_{\gamma\beta} \,, \\
  {}^{(4)}R_{\alpha\beta} &= R^{\gamma}{}_{\alpha\gamma\beta} \,, \\
  C_{\alpha\beta\gamma\delta} &= R_{\alpha\beta\gamma\delta}
                                 - \psi_{\alpha[\gamma}R_{\delta]\beta}
                                 - \psi_{\beta[\gamma}R_{\delta]\alpha}
              + \frac{1}{3}R\, \psi_{\alpha[\gamma} \psi_{\delta]\beta} \,.
\end{align}
The Ricci tensor of the spatial slices is commonly given by $R_{ab}$, so we
denote the spacetime Ricci tensor by ${}^{(4)}R_{\alpha\beta}$.

This allows us to define the Weyl scalars as follows,
\begin{align}
  \Psi_4 &= C_{\alpha\beta\gamma\delta} k^\alpha \bar{m}^\beta k^\gamma \bar{m}^\delta \,, \\
  \Psi_3 &= C_{\alpha\beta\gamma\delta} \ell^\alpha k^\beta \bar{m}^\gamma k^\delta \,, \\
  \Psi_2 &= C_{\alpha\beta\gamma\delta} \ell^\alpha m^\beta \bar{m}^\gamma k^\delta \,, \\
  \Psi_1 &= C_{\alpha\beta\gamma\delta} \ell^\alpha k^\beta \ell^\gamma m^\delta \,, \\
  \Psi_0 &= C_{\alpha\beta\gamma\delta} \ell^\alpha m^\beta \ell^\gamma m^\delta \,,
\end{align}
for a complex null tetrad given by,
\begin{align}
  \ell^\mu &= \left ( t^\mu + r^\mu \right ) / \sqrt{2} \,, \\
  k^\mu    &= \left ( t^\mu - r^\mu \right ) / \sqrt{2} \,, \\
  m^\mu    &= \left ( \theta^\mu + i\phi^\mu \right ) / \sqrt{2} \,, \label{eq:tetrad_m}
\end{align}
where $r^\mu$ is an outward pointing space-like unit vector. The orientation
of the tetrad is chosen so that in Minkowski spacetime we have
$\theta^i=\hat{x}^i$ and $\phi^i=\hat{y}^i$ on the $z$ axis and everywhere
else $\theta^i$ and $\phi^i$ are defined in the usual way on the sphere,
for more details about the tetrad see Section D in \cite{Boyle:2007ft}, but
note that they use the opposite sign defining the Weyl scalars.

The sign convention of the strain polarizations $h_{+}$ and $h_\times$
are chosen as follows. We define $h_{\mu\nu} = \psi_{\mu\nu} - \eta_{\mu\nu}$
where $\eta_{\mu\nu}$ is the Minkowski metric. For a 
gravitational wave propagating in the outward radial direction,
we define the strain as
\begin{align}
  h_{+} &= \frac{1}{2} \left(
  h_{\hat\theta\hat\theta} - h_{\hat\phi\hat\phi} \right ) \label{eq:hplus_sign} \,, \\
  h_{\times} &= h_{\hat\theta\hat\phi} \label{eq:hcross_sign} \,, \\
  h &= h_{+} - i h_{\times} \label{eq:h_from_hplus_hcross} \,,
\end{align}
where $\hat{\theta}$ and $\hat{\phi}$ correspond to the usual coordinate vectors on the sphere. From Eqs.~(\ref{eq:hplus_sign})--(\ref{eq:h_from_hplus_hcross})
we have the following relation between $\Psi_4$ and the second time
derivative of the strain,
\begin{equation} \label{eq:psi4_h}
  \lim_{r\rightarrow\infty} \Psi_4 = - \ddot{h} \,.
\end{equation}
The sign of the strain defined in Eqs.~(\ref{eq:hplus_sign})--(\ref{eq:h_from_hplus_hcross}) 
is the current definition of the strain for
the waveforms in our catalog. This differs by an overall sign for any strain
waveform previously acquired from our catalog. Please see
the end of Appendix~\ref{sec:waveform-files} for details on working with
previous and current versions of the waveform files.
In defining the Weyl scalars---and thus the relationship between the
Weyl tensor and the strain---there is a significant representation in the
literature both agreeing with our sign convention~\cite{Ashtekar:2000hw,Griffiths:2009esp,Nichols:2011pu}
and having the opposite sign convention~\cite{Boyle:2007ft, Baumgarte:2010ndz, Chandrasekhar:1983mtbh, Bishop:2016lgv}.
Newman and Penrose originally defined the Weyl scalars opposite to ours, but
they also used a $(+,-,-,-)$ metric signature~\cite{Newman:1962}.

For the Regge-Wheeler and Zerilli scalars $\Phi^{(\pm)}$ we choose
our sign convention so that
Eqs.~(\ref{eq:WaveAmplitudesFromRWZScalars})
and~(\ref{eq:strain_at_a_point}) give the same polarizations for a
linearized transverse-traceless gauge wave as Eqs.~(\ref{eq:hplus_sign})
and~(\ref{eq:hcross_sign}). This means we have the same sign of
$\Phi^{(-)}$ but the opposite sign of $\Phi^{(+)}$ as
Ref.~\cite{Rinne:2008vn}, and the same signs of $\Phi^{(\pm)}$ as
Ref.~\cite{Ruiz:2007yx}.

Our waveform quantities are decomposed in terms of spin-weighted
spherical harmonics, cf.~Eqs.~(\ref{eq:Psi4_at_a_point})
and~(\ref{eq:strain_at_a_point}).  We use the sign conventions for
spin-weighted spherical harmonics as given in Ref.~\cite{Brown2007,Boyle:2016tjj}.
In terms of $\theta$ and $\phi$ we give the spin-weight $-2$ spherical harmonics
for $l=2$ as an example,
\begin{align}
  {}_{-2}Y_{2\,\pm2} &= \sqrt{\frac{5}{64\pi}} \left( 1 \pm \cos\theta \right)^2 e^{\pm2i\phi} \,, \\
  {}_{-2}Y_{2\,\pm1} &= \sqrt{\frac{5}{16\pi}} \sin\theta \left( 1 \pm \cos\theta \right) e^{\pm i\phi} \,, \\
  {}_{-2}Y_{2\,0\hphantom{\pm}} &= \sqrt{\frac{15}{32\pi}} \sin^2\theta \,.
\end{align}

The following is a brief selection of how our sign conventions for the strain $h$
 and RWZ scalars $\Phi^{(\pm)}$ compare with a few other sources in the literature:
\begin{itemize}

  \item The signs of $h$ and $\Phi^{(+)}$ differ by an overall sign compared to our previous
    catalog release~\cite{Mroue:2013xna}.

  \item Our conventions agree with~\cite{Brown2007,Ruiz:2007yx}, except that it appears
        that Eq.~(II.5) of~\cite{Brown2007} has a sign error.

  \item Ref.~\cite{Rinne:2008vn} defines $\Phi^{(+)}$ in Eq.~(29) with the opposite sign as we use here,
        but their definition of $\Phi^{(-)}$ in Eq.~(18) agrees with ours.

  \item The same sign differences between this paper and~\cite{Rinne:2008vn} also appear
    in~\cite{Rinne:2008ig}. In~\cite{Rinne:2008ig}, Equation~(15) should have 
		an overall sign change to match our convention and the second equation (unnumbered) 
		in Sec.~3.3 should
        have the opposite sign on the $\Phi^{(+)}$ term.

  \item Ref.~\cite{Nagar:2005ea} defines vector and tensor spherical harmonics with the opposite sign
        to ours, which would indicate that their odd-parity RWZ function has the opposite sign of
        ours. However our sign for $\Phi^{(+)}$ is opposite their definition, so our overall definition
        of the strain $h \sim \Phi^{(+)} + i \Phi^{(-)}$ agrees with theirs up to a sign. This conclusion
        assumes that both papers use the same sign convention for the surface volume form $\epsilon_{AB}$,
        which is not made clear in their paper. There also appears to be a
        factor of 2 difference in the definitions of the rank-2 tensor spherical harmonics,
        but this might be due to unclear notation.%
\end{itemize}

\section*{References}
\bibliographystyle{iopart_num}
\bibliography{catalog-2017}

\end{document}